\documentclass[preprint,dvips]{elsarticle}
\usepackage{epsfig}
\usepackage{amssymb,latexsym,amsmath}
\newcommand{\eq}[1]{\begin{align} #1 \end{align}}

\newcommand{\p}{\partial}
\newcommand{\abs}[1]{\left\vert#1\right\vert}
\newcommand{\mc}[1]{\mathcal{#1}}
\newcommand{\comment}[1]{}

\begin{document}
\title{Analytical formulas for shear and bulk viscosities in relativistic gaseous mixtures with constant cross sections.}

\author{Oleg Moroz \footnote{Tel.: +38 044 5213458}}
\ead{ moroz@bitp.kiev.ua }
\address{Bogolyubov Institute for Theoretical Physics, 14-b, Metrolohichna str., Kyiv, 03680, Ukraine}

\date {\today}

\begin{abstract}
Using Mathematica package, we derive analytical closed-form expressions for the shear and the bulk viscosity
coefficients in multicomponent relativistic gases with constant cross sections, being the relativistic
generalization for the hard spheres model. Some of them are cumbersome and require symbolic manipulations in an
algebraic package. The constant cross sections are of the elastic processes, while the inelastic (or
number-changing) processes (collisions or decays) are considered only partly. As examples, we find explicit
expressions of the correct single-component first-order shear viscosity coefficient and some explicit analytical
results for the binary mixture. These formulas have numerous applications in approximate nonequilibrium
descriptions of gases of particles or quasiparticles with averaged (temperature dependent) cross sections. In
addition to this, we present formulas for collision rates and some other related formulas. This paper is
a shortened version of a previous one.
\end{abstract}

\begin{keyword}
bulk viscosity \sep shear viscosity \sep kinetics \sep transport theory

\PACS 47.45.Ab \sep 51.20.+d \sep 05.20.Dd

\end{keyword}

\maketitle

\section{ Introduction }
The bulk and the shear viscosity coefficients are transport coefficients which enter the hydrodynamic equations,
and, thus, are important for studying of nonequilibrium evolution of any thermodynamic system. In this regard,
another way of dissipative nonequilibrium description can be mentioned \cite{dissipcoef, dissipcoef2}. In
rarefied gases of particles or quasiparticles with short-range interactions the viscosities can be calculated
in a perturbative regime\footnote{In non-abelian gauge theories there are also contributions from non-perturbative effects \cite{Arnold:2002zm}.}.
The leading contribution can be obtained in the framework of the Boltzmann equation (BE),
derivable within the BBGKY hierarchy with the well-known assumption for correlations \cite{landau10} (Sec. 16).
The BE's justification from first principles and the next-to-leading order corrections have been obtained in
calculations for weakly coupled quantum field theories by the Kubo (or Green-Kubo\footnote{The Kubo formulas are distinguished from the
Green-Kubo formulas, e. g., in \cite{Kadanoff, Muronga:2003tb}.}) formulas \cite{jeon, Gagnon:2006hi, Gagnon:2007qt, jeonyaffe, Arnold:2002zm}\footnote{
There is, however, a special important issue connected with particle number conservation/nonconservation for the bulk viscosity \cite{Moroz:2013vd}.}.

The aim of this paper is to derive analytical closed-form (through special and/or elementary functions)
expressions for the shear and the bulk viscosities in multicomponent relativistic gases with constant cross
sections, and similarly for the collision rates and some other related quantities. Previously the
single-component viscosities were obtained in \cite{anderson} (we correct the shear viscosity), being the
relativistic generalization of the ones in the hard spheres model \cite{landau10} (Secs. 8, 10).

The structure of the paper is the following. Sec. \ref{CalcSec} contains some comments on cross sections,
connection with the hard spheres model, most of the used designations and methodology. In Sec. \ref{ResSec} we
present explicit analytical results for the single-component gas, including the ones for the nonequilibrium
distribution function (DF) and some analysis for the inelastic processes. We also present some explicit
expressions for the binary mixture and the collision rates (and related quantities) in the multicomponent
mixture. In Sec. \ref{DiscSec} we discuss significance and applications of the obtained formulas. In Sec.
\ref{ConcSec} we state the conclusions. Transformations of collision brackets, being the 12-dimensional
integrals, which enter the viscosities, and some analytical formulas for them can be found in \ref{appJ}.

This paper is a shortened version of the previous one \cite{Moroz:2011vn}.

\section{ Methodology \label{CalcSec} }

\subsection{ Some comments on cross sections and effective radiuses \label{hardcorsec}}

For compatibility with previously obtained results and from practical considerations we want to introduce an
effective radius $r$ through the hard core repulsion model or the model of hard spheres. The differential
scattering cross section for this model can be inferred from the problem of scattering of point particle on the
spherical potential ${U(r)=\infty}$ if ${r\leq a}$ and ${U(r)=0}$ if ${r>a}$ \cite{landau1}. In this model the
differential cross section is equal to $a^2/4$. To apply this result to the gas of hard spheres with the radius
$r$ one can notice that the scattering of any two spheres can be considered as the scattering of the point
particle on the sphere of the radius $2r$, so that one should take ${a=2r}$. The total cross section
$\sigma_{tot}$ is obtained after integration over the angles of the $r^2 d\Omega$, which results in the
$\sigma_{tot}=4\pi r^2$. For collisions of hard spheres of different radiuses one should take ${a=r_k+r_l}$ or
replace the $r$ on the $\frac{r_k+r_l}2$:
 \eq{\label{hccs}
 \sigma_{tot,kl}=\pi (r_k+r_l)^2.
 }
The relativistic generalization of this model is the constant (not dependent on the scattering energy and angle)
differential cross sections model.

The hard spheres model is classical, and connection of its cross sections to cross sections, calculated in any
quantum theory, is needed. For particles, having a spin, the differential cross sections averaged over the
initial spin states and summed over the final ones will be used\footnote{It's assumed that particle numbers of
the same species but with different spin states are equal. If this were not so, then in approximation in which
the spin interactions are neglected and probabilities to have certain spin states are equal the numbers of the
particles with different spin states would be approximately equal in the mean free time. With equal particle
numbers their DFs are equal too. This allows one to use the summed over the final states cross sections in the
BEs.}. If colliding particles are identical and their differential cross section is integrated over the momentums
(or the spatial angle to get the total cross section) then it should be multiplied on the factor $\frac12$ to
cancel double counting of the momentum states. These factors are exactly the factors $\gamma_{kl}$ next to the
collision integrals in the BEs (\ref{boleqs}). The differential cross sections times these factors will be called
(adopting the terminology mentioned in \cite{groot}) the classical differential cross sections.

\subsection{ The system of the BEs and its solution \label{CalcSecA} }

The methodology in this paper goes close to the ones in the monograph \cite{groot}, though with some differences
(including corrections of a couple of typos) and generalizations. We find it very instructive to include
compilation of some pieces of the methodology (some of which are absent in the \cite{groot}) for convenience and
completeness, adding some comments and tacitly answering some questions. We use units $\hbar=c=k_B=1$ throughout
the paper by default. Conversion factors can be found, e. g., in \cite{units}. Let's start from some definitions.
We use the designations of the \cite{groot} mostly. Multi-indices $k,l,m,n$ will be used to denote particle
species with certain spin states. Indexes $k',l',m',n'$ will be used to denote particle species without regard to
their spin states (and run from 1 to the number of the particle species $N'$) and $a,b$ to denote conserved
quantum numbers\footnote{In systems with only the elastic collisions each particle species have their own
"conserved quantum number", equal to 1.}. Quantifiers $\forall$ with respect to the indexes are omitted in the
text where they may be needed, which won't result in a confusion. As nothing depends on spin variables, one has
for every sum over the multi-indexes
 \eq{
  \sum_k ... = \sum_{k'}g_{k'}...,
 }
where $g_{k'}$ is the spin degeneracy factor. The following
assignments will be used:
 \begin{eqnarray}\label{assign1}
  \nonumber n &\equiv& \sum_{k}n_k \equiv \sum_{k'} n_{k'}, \quad n_a\equiv \sum_k q_{ak} n_k,
  \quad x_k\equiv\frac{n_k}{n}, \quad x_a\equiv\frac{n_a}{n}, \\
  \hat \mu_k &\equiv& \frac{\mu_k}{T}, \quad \hat \mu_a \equiv \frac{\mu_a}{T},
  \quad z_k\equiv\frac{m_k}{T}, \quad \pi_k^\mu\equiv \frac{p_k^\mu}{T},
  \quad \tau_k\equiv \frac{p_k^\mu U_\mu}{T},
 \end{eqnarray}
where $q_{ak}$ denotes values of conserved quantum numbers of the $a$-th kind of the $k$-th particle species.
Everywhere the particle number densities are summed the spin degeneracy factor $g_{k'}$ appears and then gets
absorbed into the $n_{k'}$ or the $x_{k'}$ by the definition. All other quantities with primed and unprimed
indexes don't differ, except for rates, the mean free times and the mean free paths defined in Sec.
\ref{appmfp}, the $\gamma_{kl}$ commented below, the coefficients $A_{k'l'}^{rs}$, $C_{k'l'}^{rs}$ and, of
course, quantities, whose free indexes set the indexes of the particle number densities $n_k$. The assignment
${\int \frac{d^3p_k}{p_k^0}\equiv \int_{p_k}}$ will also be used for compactness somewhere.

The particle number flows are\footnote{The $+,-,-,-$ metric
signature is used throughout the paper.}
 \eq{\label{pflow}
 N^{\mu}_k=\int \frac{d^3p_k}{(2\pi)^3p^0_k} p^\mu_k f_k,
 }
where the assignment ${f_k(p_k)\equiv f_k}$ is introduced. The
energy-momentum tensor is
 \eq{\label{enmomten}
 T^{\mu\nu}=\sum_k \int \frac{d^3p_k}{(2\pi)^3p_k^0}p_k^\mu p_k^\nu f_k.
 }
The local equilibrium DFs are
 \eq{\label{loceq}
 f^{(0)}_k=e^{(\mu_k-p_k^\mu U_\mu)/T},
 }
where $\mu_k$ is the chemical potential of the $k$-th particle species, $T$ is the temperature and $U_\mu$ is the
relativistic flow 4-velocity such that ${U_\mu U^\mu=1}$ (with a frequently used consequence ${U_\mu\p_\nu
U^\mu=0}$). The local equilibrium implies perturbation of the independent thermodynamic variables and the flow
velocity over a \emph{global} equilibrium (see just below) such that they can depend on the space-time coordinate
$x^\mu$. We assume chemical equilibrium, which implies that the particle number densities are equal to their
global equilibrium values. We call the global equilibrium as the time-independent stationary state with the
maximal entropy\footnote{The kinetic equilibrium implies that the momentum distributions are the same as in the
global equilibrium. Thus, a state of a system with both the pointwise (for the whole system) kinetic and the
pointwise chemical equilibria is the global equilibrium.}. The global equilibrium state of an isolated system can
be found by variation of the total nonequilibrium entropy functional \cite{landau5} (Sec. 40) over the DFs with
condition of the total energy and the total net charges conservation:
\begin{eqnarray}\label{Uf}
 \nonumber U[f] &=& \sum_k \int \frac{d^3p_kd^3x}{(2\pi)^3}
 f_k(1-\ln f_k)-\sum_k\int \frac{d^3p_kd^3x}{(2\pi)^3}\beta p^0_kf_k\\
 &-&\sum_{a,k} \lambda_a q_{ak} \int \frac{d^3p_kd^3x}{(2\pi)^3} f_k,
\end{eqnarray}
where $\beta, \lambda_a$ are the Lagrange coefficients. Equating the first variation to zero, one easily gets the
function (\ref{loceq}) with ${U^\mu=(1,0,0,0)}$, ${\beta=\frac1{T}}$ and
 \eq{\label{mukdef}
 \mu_k=\sum_a q_{ak}\mu_a,
 }
where ${\mu_a=\lambda_a}$ are the independent chemical potentials
coupled to the conserved net charges.

With ${f_k=f_k^{(0)}}$, substituted in the (\ref{pflow}) and the
(\ref{enmomten}), one gets the leading contribution in the
gradients expansion of the particle number flow and the
energy-momentum tensor:
 \eq{
 N^{(0)\mu}_k=n_kU^\mu,
 }
 \eq{\label{T0}
 T^{(0)\mu\nu}=\epsilon U^\mu U^\nu - P\Delta^{\mu\nu},
 }
where the projector
 \eq{\label{proj}
 \Delta^{\mu\nu}\equiv g^{\mu\nu}-U^\mu U^\nu,
 }
is introduced. Above $n_k$ is the ideal gas (IG) particle number density, $\epsilon$ is the IG energy density,
$P$ is the IG pressure \cite{groot} (Chap. II, Sec. 4). Also, the following assignments are used:
 \begin{eqnarray}\label{assign2}
  e&\equiv&\frac{\epsilon}{n}=\sum_kx_ke_k, \quad h_k\equiv e_k+T,
  \quad h\equiv\frac{\epsilon+P}{n}=\sum_k x_kh_k, \\
  \nonumber \hat e_k&\equiv&\frac{e_k}{T}=z_k\frac{K_3(z_k)}{K_2(z_k)}-1,
  \quad \hat e\equiv\frac{e}{T}, \quad \hat h_k\equiv\frac{h_k}{T}=
  z_k\frac{K_3(z_k)}{K_2(z_k)}, \quad \hat h\equiv \frac{h}{T}.
 \end{eqnarray}
Above $h$ is the enthalpy per particle, $e$ is the energy per particle and $h_k$, $e_k$ are the enthalpy and the
energy per particle of the $k$-th particle species correspondingly, which are well defined in the IG.

In the relativistic hydrodynamics the flow velocity, $U^\mu$, needs somewhat extended definition. A convenient
condition which can be applied to the $U^\mu$ is the Landau-Lifshitz condition \cite{landau6} (Sec. 136). This
condition states that in the local rest frame (where the flow velocity is zero though its gradient can have a
nonzero value) each imaginary infinitesimal cell of fluid should have zero momentum, and its energy density and
the charge density should be related to other thermodynamic quantities through the equilibrium thermodynamic
relations (without a contribution of nonequilibrium dissipations). Its covariant mathematical formulation is
 \eq{\label{lLcond}
 (T^{\mu\nu}-T^{(0)\mu\nu})U_\mu=0, \quad (N^\mu_a-N^{(0)\mu}_a)U_\mu=0.
 }
The next to leading correction over the gradients expansion to the $T^{\mu\nu}$ can be written as an expansion
over the 1-st order Lorentz covariant gradients, which are rotationally and space inversion invariant and satisfy
the Landau-Lifshitz condition\footnote{ This form of $T^{(1)\mu\nu}$ also respects the second law of
thermodynamics \cite{landau6} (Sec. 136). } (\ref{lLcond}):
\begin{eqnarray}\label{T1}
 T^{(1)\mu\nu} &\equiv& 2\eta \overset{\circ}{\overline{\nabla^\mu
 U^\nu}}+\xi \Delta^{\mu\nu} \nabla_\rho U^\rho \\
 \nonumber &=& \eta\left(\Delta^\mu_\rho \Delta^\nu_\tau
 +\Delta^\nu_\rho \Delta^\mu_\tau-\frac23\Delta^{\mu\nu}\Delta_{\rho\tau}\right)\nabla^\rho
 U^\tau+\xi \Delta^{\mu\nu} \nabla_\rho U^\rho,
\end{eqnarray}
where for any tensor $a_{\mu\nu}$ the symmetrized traceless tensor assignment is introduced:
\begin{eqnarray}\label{tracelessten}
 \overset{\circ}{\overline{a_{\mu\nu}}} &\equiv& \left(\frac{\Delta_{\mu\rho}
 \Delta_{\nu\tau}+\Delta_{\nu\rho} \Delta_{\mu\tau}}2-\frac13\Delta_{\mu\nu}
 \Delta_{\rho\tau}\right)a^{\rho\tau}\equiv \Delta_{\mu\nu\rho\tau}a^{\rho\tau}, \\
 \nonumber &~& \Delta^{\mu\nu}_{~~\rho\tau}\Delta^{\rho\tau}_{~~\sigma\lambda}=
 \Delta^{\mu\nu}_{~~\sigma\lambda}.
\end{eqnarray}
The equation (\ref{T1}) is the definition of the shear $\eta$ and the bulk $\xi$ viscosity coefficients. The $\xi
\Delta^{\mu\nu} \nabla_\rho U^\rho$ term in the (\ref{T1}) can be considered as a nonequilibrium contribution to
the pressure, entering the (\ref{T0}).

By means of the projector (\ref{proj}) one can split the
space-time derivative $\p_\mu$ as
 \eq{
 \p_\mu=U_\mu U^\nu \p_\nu + \Delta_\mu^\nu\p_\nu = U_\mu
 D+\nabla_\mu,
 }
where $D \equiv U^\nu \p_\nu$, $\nabla_\mu \equiv \Delta_\mu^\nu\p_\nu$. In the local rest frame (where
${U^\mu=(1,0,0,0)}$) the $D$ becomes the time derivative and the $\nabla_\mu$ becomes the spacial derivative.
Then the BEs can be written in the form
 \eq{\label{boleqs}
 p_k^\mu\p_\mu f_k=(p_k^\mu U_\mu D + p_k^\mu \nabla_\mu )f_k
 =C_k^{el}[f_k]+C_k^{inel}[f_k],
 }
where $C_k^{inel}[f_k]$ represents the inelastic collision integrals (it is omitted in calculations in this paper
if the opposite is not stated explicitly) and $C_k^{el}[f_k]$ is the elastic ${2\leftrightarrow2}$ collision
integral. The collision integral $C_k^{el}[f_k]$ has the form of the sum of positive gain terms and negative loss
terms. Its explicit form is\footnote{The factor $\gamma_{kl}$ cancels double counting in integration over
momentums of identical particles. The factor $\frac12$ comes from the relativistic normalization of the
scattering amplitudes. } (cf. \cite{jeon, Arnold:2002zm})
 \begin{eqnarray}\label{ckel}
  \nonumber C_k^{el}[f_k]&=&\sum_{l} \gamma_{kl}\frac12\int
  \frac{d^3p_{1l}}{(2\pi)^32p_{1l}^0}\frac{d^3p'_k}{(2\pi)^32{p'}_k^0}
  \frac{d^3p'_{1l}}{(2\pi)^32{p'}_{1l}^0}(f'_{k}f'_{1l}-f_{k}f_{1l})\\
  &\times& |\mc M_{kl}|^2(2\pi)^4\delta^{4}(p'_k+p'_{1l}-p_k-p_{1l}),
 \end{eqnarray}
where ${\gamma_{kl}=\frac12}$ if $k$ and $l$ denote the same
particle species without regard to the spin states and
${\gamma_{kl}=1}$ otherwise, ${|\mc M_{kl} (p'_k,p'_{1l};
p_k,p_{1l})|^2 \equiv |\mc M_{kl}|^2}$ is the square of the
dimensionless elastic scattering amplitude averaged over the
initial spin states and summed over the final ones. Index $1$
designates that $p_k$ and $p_{1k}$ are different variables.
Introducing ${W_{kl}\equiv W_{kl}(p'_k,p'_{1l};p_k,p_{1l})}$ as
 \eq{
 W_{kl}=\frac{|\mc M_{kl}|^2}{64\pi^2}\delta^{4}(p'_k+p'_{1l}-p_k-p_{1l}),
 }
one can rewrite the collision integral (\ref{ckel}) in the form as
in \cite{groot} (Chap. I, Sec. 2)
 \eq{\label{ckelgroot}
 C_k^{el}[f_k]=(2\pi)^3\sum_{l}\gamma_{kl}
 \int_{p_{1l},{p'}_k,{p'}_{1l}}\left(\frac{f'_{k}}{(2\pi)^3}\frac{f'_{1l}}{(2\pi)^3}
 -\frac{f_{k}}{(2\pi)^3}\frac{f_{1l}}{(2\pi)^3}\right)W_{kl}.
 }
The $W_{kl}$ is related to the elastic differential cross section
$\sigma_{kl}$ as \cite{groot} (Chap. I, Sec. 2)
 \eq{
 W_{kl}=s\sigma_{kl}\delta^{4}(p'_k+p'_{1l}-p_k-p_{1l}),
 }
where ${s=(p_k+p_{1l})^2}$ is the usual Mandelstam variable. The
$W_{kl}$ has properties $W_{kl}(p'_k,p'_{1l};p_k,p_{1l}) =
W_{kl}(p_k,p_{1l};p'_{k},p'_{1l}) =
W_{lk}(p'_{1l},p'_{k};p_{1l},p_k)$ (due to time reversibility and
a freedom of relabelling of order numbers of particles taking part
in reaction). And, e. g., $W_{kl}(p'_k,p'_{1l};p_k,p_{1l}) \neq
W_{kl}(p'_{1l},p'_{k};p_{1l},p_{k})$ in the general case. The
elastic collision integrals have important properties which one
can easily prove \cite{groot} (Chap. II, Sec. 1):
 \eq{\label{c22prop0}
 \int \frac{d^3p_k}{(2\pi)^3p^0_k} C_k^{el}[f_k]=0,
 }
 \eq{\label{c22prop}
 \sum_k\int \frac{d^3p_k}{(2\pi)^3p^0_k} p_k^\mu C_k^{el}[f_k]=0.
 }
Also, the $C^{el}_k[f_k]$ vanishes if ${f_k=f^{(0)}_k}$.

The DFs $f_k$ solving the system of the BEs approximately are sought in the form (below the essence of the
Chapman-Enskog method is reproduced, see, e. g., \cite{groot}, Chap. V)
 \eq{\label{fpert}
 f_k=f^{(0)}_k+f^{(1)}_k\equiv f^{(0)}_k+f^{(0)}_k\varphi_k(x,p_k),
 }
where it's assumed that $f_k$ depend on the $x^\mu$ entirely through the $T$, $\mu_k$, $U^\mu$ or their
space-time derivatives. It is also assumed that ${|\varphi_k|\ll 1}$. After substitution of ${f_k=f^{(0)}_k}$ in
the (\ref{boleqs}) the r. h. s. becomes zero and the l. h. s. is zero only if the $T$, $\mu_k$ and $U^\mu$ don't
depend on the $x^\mu$ (provided they don't depend on the momentum $p^\mu_k$). The 1-st order space-time
derivatives of the $T$, $\mu_k$, $U^\mu$ in the l. h. s. should be cancelled by the first nonvanishing
contribution in the r. h. s. This means that the $\varphi_k$ should be proportional to the 1-st order space-time
derivatives of the $T$, $\mu_k$, $U^\mu$. The covariant time derivatives $D$ can be expressed through the
covariant spacial derivatives by means of approximate hydrodynamic equations, valid at the same order in the
gradients expansion. Let's derive them. Integrating the (\ref{boleqs}) over the $\frac{d^3p_k}{(2\pi)^3p^0_k}$
with the ${f_k=f^{(0)}_k}$ in the l. h. s. with the inelastic collision integrals retained and using the
(\ref{c22prop0}) and the (\ref{pflow}) one would get (which can be justified using explicit form of the inelastic
collision integrals)
 \eq{\label{conteq}
 \p_\mu N^{(0)\mu}_k=Dn_k+n_k\nabla_\mu U^\mu=I_k,
 }
where $I_k$ is the sum of the inelastic collision integrals integrated over the momentum. It is responsible for
the nonconservation of the total particle number of the $k$-th particle species and has the property ${\sum_k
q_{ak} I_k=0}$. If ${C^{inel}_k[f_k]=0}$, then ${I_k=0}$ which results in conservation of the total particle
numbers of each particle species. Multiplying the (\ref{conteq}) on the $q_{ak}$ and summing over $k$ one gets
the continuity equations for the net charge flows (cf. \cite{groot}, Chap. II, Sec. 1):
 \eq{\label{conteq2}
 \p_\mu N^{(0)\mu}_a=Dn_a+n_a\nabla_\mu U^\mu=0.
 }
Also, integrating the (\ref{boleqs}) over the $p_k^\mu\frac{d^3p_k}{(2\pi)^3p^0_k}$ with the ${f_k=f^{(0)}_k}$ in
the l. h. s., one gets
 \eq{\label{encons0}
 \p_\rho T^{(0)\rho\nu}=\p_\rho(\epsilon U^\rho U^\nu-P\Delta^{\rho\nu})=0.
 }
There is zero in the r. h. s. even if the inelastic collision integrals are retained because they respect energy
conservation too. Note that the BEs (\ref{boleqs}) (without any thermal corrections) permit a self-consistent
description only if the energy-momentum tensor and the net charge flows of the IG are used. After the convolution
of the (\ref{encons0}) with the $\Delta^\mu_\nu$ one gets the Euler's equation:
 \eq{\label{eulereq}
 DU^\mu=\frac1{\epsilon+P}\nabla^\mu P=\frac1{hn}\nabla^\mu P.
 }
After the convolution of the (\ref{encons0}) with the $U_\nu$ one
gets equation for the energy density:
 \eq{\label{encons1}
 D\epsilon=-(\epsilon+P)\nabla_\mu U^\mu = - hn\nabla_\mu U^\mu.
 }

To proceed farther one needs to expand the l. h. s. of the BEs (\ref{boleqs}) over the gradients of thermodynamic
variables and the flow velocity. Let's choose the $\mu_a$ and the $T$ as the independent thermodynamic variables.
Then for the $Df^{(0)}_k$ one can write the expansion
 \eq{\label{Dfk}
 Df^{(0)}_k=\sum_a \frac{\p f^{(0)}_k}{\p\mu_a}D\mu_a+\frac{\p f^{(0)}_k}{\p
 T}DT+\frac{\p f^{(0)}_k}{\p U^\mu}DU^\mu.
 }
Writing the expansion for the $Dn_a$ and the $D\epsilon$ one gets
from the (\ref{conteq2}) and the (\ref{encons1}):
 \eq{\label{Dna}
 Dn_a=\sum_b \frac{\p n_a}{\p \mu_b}D\mu_b+\frac{\p n_a}{\p T}DT
 =-n_a\nabla_\mu U^\mu,
 }
 \eq{\label{Depsilon}
 D\epsilon=\frac{\p\epsilon}{\p T}DT+\sum_a\frac{\p \epsilon}{\p \mu_a}D\mu_a
 =-hn\nabla_\mu U^\mu.
 }
The solution to the system of equations (\ref{Dna}),
(\ref{Depsilon}) can be found easily:
 \eq{\label{Teqn}
 DT=-R T\nabla_\mu U^\mu,
 }
 \eq{\label{mueqn}
 D\mu_a=T\sum_b \tilde{A}^{-1}_{ab}(R B_b-x_b)\nabla_\mu U^\mu,
 }
where
 \eq{\label{Rdef}
 R\equiv\frac{\hat h-\sum_{a,b} E_a \tilde{A}^{-1}_{ab}x_b}{C_{\{\mu\}}-\sum_{a,b}
 E_a\tilde{A}^{-1}_{ab}B_b},
 }
and
 \eq{
 \frac{\p n_a}{\p \mu_b}\equiv \frac{n}{T}\tilde{A}_{ab},
 \quad \frac{\p n_a}{\p T}\equiv \frac{n}{T}B_a, \quad
 \frac{\p\epsilon}{\p T}\equiv n C_{\{\mu\}},\quad
 \frac{\p\epsilon}{\p\mu_a}\equiv n E_a.
 }
Above it is assumed that the matrix $\tilde{A}_{ab}$ is not degenerate\footnote{One can prove that the $N''
\times N''$ matrix $\tilde{A}_{ab}$ in (\ref{ABCE}) is not degenerate if there are $N''$ linearly independent
conserved charges. Then one can prove that the denominator in the (\ref{Rdef2}) is not zero.}, which is related
to the self-consistency of the statistical description of the system. Using the $n_k$ and $\epsilon$ IG formulas
one gets
 \begin{eqnarray}\label{ABCE}
  \nonumber \tilde A_{ab} &=& \sum_k q_{ak}q_{bk}x_k, \quad
  E_a=\sum_k q_{ak} x_k \hat e_k, \quad B_a=E_a-\sum_b \tilde A_{ab}\hat\mu_b,\\
  \nonumber C_{\{\mu\}}&=&\sum_k x_k(3\hat h_k+z_k^2-\hat \mu_k\hat
  e_k)=\sum_kx_k(3\hat h_k+z_k^2)-\sum_aE_a\hat\mu_a \\
  &\equiv& \widetilde C_{\{\mu\}}-\sum_aE_a\hat\mu_a,
 \end{eqnarray}
and simplified expressions for the $R$ and the $D\hat\mu_a$
 \eq{\label{Rdef2}
 R=\frac{\hat h-\sum_{a,b} E_a \tilde{A}^{-1}_{ab}x_b}{\widetilde C_{\{\mu\}}-\sum_{a,b}
 E_a\tilde{A}^{-1}_{ab}E_b},
 }
 \eq{
 D\hat\mu_a=\sum_b\tilde{A}^{-1}_{ab}(R E_b-x_b)\nabla_\mu U^\mu.
 }
For the special case of the vanishing chemical potentials,
$\mu_a\rightarrow 0$, (for a chargeless system the result is the
same) the quantities $n_a$, $x_a$, $B_a$, $E_a$ tend to zero
because the contributions from particles and anti-particles cancel
each other and the chargeless particles don't contribute. Then
from the (\ref{Teqn}) and the (\ref{mueqn}) one finds
 \eq{\label{Teqnmu0}
 DT|_{\mu_a=0}=-\frac{h}{\widetilde C_{\{\mu\}}}\nabla_\mu U^\mu, \quad D\mu_a|_{\mu_a=0}=0.
 }
So if there is an anti-particle for each charged particle (which is so for the exactly conserved charges), vanishing chemical
potentials are equivalent to their exclusion from the DFs provided the spacial derivatives of the chemical potentials can be excluded.
In systems with only the elastic collisions each particle has its own
charge, so that one takes ${q_{ak}=\delta_{ak}}$ and gets
 \begin{eqnarray}\label{elquant}
  \tilde A_{kl}&=&\delta_{kl}x_k, \quad B_k=x_k(\hat e_k-\hat \mu_k),
  \quad E_k=\hat e_k x_k, \quad R=\frac1{c_\upsilon}, \\
  \nonumber C_{\{\mu\}}&-&\sum_{a,b}E_a \tilde A^{-1}_{ab} B_{b}=\sum_kx_k(-\hat h_k^2+5\hat
  h_k+z_k^2-1)\equiv\sum_kx_k c_{\upsilon,k}\equiv c_\upsilon.
 \end{eqnarray}
Then, the equation for the $DT$ (\ref{Teqn}) remains the same with a new $R$ from the (\ref{elquant}), and the
equations (\ref{mueqn}) become
 \eq{\label{mukeqn}
 D\mu_k=\left(\frac{T}{c_\upsilon}(\hat e_k-\hat \mu_k)-T\right)\nabla_\mu U^\mu.
 }
Note that in systems with only the elastic collisions the $D\mu_k$ does not tend to zero for the vanishing
chemical potentials so that the $\mu_k$ could not be omitted in the DFs in this case. As the heat conductivity
and diffusion are not considered in this paper their nonequilibrium gradients are taken equal to zero,
$\nabla_\nu P=\nabla_\nu T=\nabla_\nu\mu_a=0$. Using the (\ref{Teqn}), (\ref{mueqn}) and (\ref{eulereq}) the l.
h. s. of the (\ref{boleqs}) can be transformed as
 \eq{\label{boleqnlhs}
 (p_k^\mu U_\mu D+p_k^\mu\nabla_\mu)f_k^{(0)} = -Tf_k^{(0)}
 \pi_k^\mu \pi_k^\nu \overset{\circ}{\overline{\nabla_\mu
 U_\nu}}+Tf_k^{(0)}\hat Q_k\nabla_\rho U^\rho,
 }
where
 \eq{\label{Qsource}
 \hat Q_k\equiv\tau_k^2\left(\frac13-R\right)+\tau_k\sum_{a,b}q_{ak}
 \tilde A^{-1}_{ab}(RE_b-x_b)-\frac13z_k^2.
 }
The $\hat Q_k$ is of a universal and convenient form, see Sec. \ref{DiscSec} for discussions.
Using the (\ref{tracelessten}) one can notice that the useful (tacitly used) equality $\pi_k^\mu \pi_k^\nu
\overset{\circ}{ \overline{\nabla_\mu U_\nu} } = \overset{\circ}{ \overline{ \pi_k^\mu \pi_k^\nu }
}\overset{\circ}{ \overline{ \nabla_\mu U_\nu } }$ holds. In systems with only the elastic collisions the $\hat
Q_k$ simplifies in agreement with \cite{groot} (Chap. V, Sec. 1):
 \eq{\label{Qsource2}
 \hat Q_k=\left(\frac43-\gamma\right)\tau_k^2+
 \tau_k((\gamma-1)\hat h_k-\gamma)-\frac13z_k^2.
 }
The $\gamma$ from the \cite{groot} can be expressed through the $c_\upsilon$ (\ref{elquant}) as ${\gamma\equiv
\frac1{c_\upsilon}+1}$. The approximate solution of the BEs, which we use in what follows, is connected with
inner product, denoted as
 \eq{
 (F,G)_k\equiv\frac1{4\pi z_k^2K_2(z_k)T^2}\int_{p_k} F(p_k)G(p_k)e^{-\tau_k}.
 }
Also, we use assignments
 \eq{\label{algama}
 \alpha_k^r\equiv(\hat Q_k,\tau_k^r)_k, \quad \gamma_k^r\equiv(\tau_k^r
 \overset{\circ}{\overline{\pi_k^\mu\pi_k^\nu}},
 \overset{\circ}{\overline{\pi_{k\mu}\pi_{k\nu}}})_k, \quad
 a^r_k\equiv(1,\tau_k^r)_k.
 }
Expressions of the $\alpha_k^r$ and the $\gamma_k^r$ through the $a_k^r$, the recurrence relations for the
$a_k^r$, some explicit expressions for the $\alpha_k^r$, $\gamma_k^r$ and $a_k^r$ can be found in the
\cite{groot} (Chap. VI, Sec. 1, App.). Using the latter ones, we find for the quantities of a special interest
$\alpha_k^0$ and $\alpha_k^1$ in systems with elastic and inelastic processes
 \eq{\label{alphak0}
 \alpha_k^0=1+\sum_{a,b}q_{ak} \tilde A^{-1}_{ab}(R E_b-x_b)-\hat e_k R,
 }
 \eq{\label{alphak1}
 \alpha_k^1=\hat h_k+\sum_{a,b}\hat e_k q_{ak}\tilde A^{-1}_{ab}(R E_b-x_b)
 -(3\hat h_k+z_k^2)R.
 }
The $\alpha_k^0$ and the $\alpha_k^1$ should satisfy the equations \cite{groot} (Chap. VI, Sec. 3)
 \eq{\label{lhsnchcons}
 \sum_kq_{ak}x_k\alpha_k^0=0,
 }
 \eq{\label{lhsencons}
 \sum_kx_k\alpha_k^1=0,
 }
(for consistency) which can be explicitly checked using the (\ref{alphak0}) and the (\ref{alphak1}). In the
partial case of only elastic processes one gets \cite{groot} (Chap. VI, Sec. 3)
 \eq{\label{elal0}
 \alpha_k^0=0.
 }
The (\ref{lhsnchcons}) and the (\ref{lhsencons}) also have relation to the local conservation laws, being
warranted by the conservations laws (\ref{conteq2}) and (\ref{encons0}). Analogical quantities
$(1,\overset{\circ}{\overline{\pi^\mu_{k}\pi^\nu_{k}}})$ and
$(p_k^\lambda,\overset{\circ}{\overline{\pi^\mu_{k}\pi^\nu_{k}}})$ vanish automatically because of the special
tensorial structure\footnote{ Direct computation gives $(1,\overset{\circ}{ \overline{\pi^\mu_k \pi^\nu_k} })_k
\propto (C_1 U^\sigma U^\rho + C_2\Delta^{\sigma\rho}) \Delta_{ ~~\sigma\rho }^{\mu\nu} = 0$, $(p_k^\lambda,
\overset{\circ}{ \overline{\pi^\mu_k \pi^\nu_k} })_k \propto (C_1 U^\lambda U^\sigma U^\rho + C_2U^\lambda
\Delta^{\sigma\rho} + C_3U^\sigma \Delta^{\lambda\rho}) \Delta_{~~\sigma\rho}^{\mu\nu} = 0$.} of the
$\overset{\circ}{ \overline{ \pi^\mu_{k} \pi^\nu_{k}} }$.

The next step is to transform the r. h. s. of the BEs (\ref{boleqs}). After the substitution of the (\ref{fpert})
in the r. h. s. of the (\ref{boleqs}) the collision integrals become linear, and one gets
 \eq{\label{boleqnrhs}
 C_k^{el}[f_k]\approx -f_k^{(0)}\sum_l \mc L_{kl}^{el}[\varphi_k],
 }
where
 \eq{
 \mc L_{kl}^{el}[\varphi_k]\equiv\frac{\gamma_{kl}}{(2\pi)^3}\int_{p_{1l},{p'}_k,{p'}_{1l}}
 f_{1l}^{(0)}(\varphi_k+\varphi_{1l}-\varphi'_k-\varphi'_{1l})W_{kl}.
 }
The unknown functions $\varphi_k$ are sought in the form
 \eq{\label{varphi}
 \varphi_k=\frac1{n\sigma(T)}\left(-A_k(p_k)\nabla_\mu U^\mu+C_k(p_k)
 \overset{\circ}{\overline{\pi^\mu_k \pi^\nu_k}}
 \overset{\circ}{\overline{\nabla_\mu U_\nu}}\right),
 }
where $\sigma(T)$ is some formal averaged cross section, used to come to dimensionless quantities. Then, using
the (\ref{boleqnlhs}) and the (\ref{boleqnrhs}), and the fact that the gradients $\nabla_\mu U^\mu$ and
$\overset{\circ}{\overline{\nabla_\mu U_\nu}}$ are independent, from the BEs independent integral equations
follow:
 \eq{\label{xieqn}
 \hat Q_k=\sum_l x_l L_{kl}^{el}[A_k],
 }
 \eq{\label{etaeqn}
 \overset{\circ}{\overline{\pi_{k}^\mu \pi_{k}^\nu}}=
 \sum_l x_l L_{kl}^{el}[C_k \overset{\circ}{\overline{\pi_{k}^\mu \pi_{k}^\nu}}],
 }
where the dimensionless collision integrals are introduced:
 \eq{
 L_{kl}^{el}[\chi_k]=\frac1{n_lT\sigma(T)}\mc L_{kl}^{el}[\chi_k].
 }
In the case of present inelastic processes the l. h. s. of the (\ref{xieqn}) is set by the source term
(\ref{Qsource}) and the r. h. s. contains the linear inelastic collision integrals. After introduction of
inelastic processes the source terms in the (\ref{xieqn}) become much larger as demonstrated in Sec.
\ref{singcomsec}. Using the equations (\ref{mueqn}) and (\ref{Teqn}) and the IG formulas (\ref{ABCE}) one can
check that in the zero masses limit the source terms $\hat Q_k$ (\ref{Qsource}) tend to zero (eventually
resulting into zero bulk viscosity in the considered approximation) and ${D\hat\mu_a=0}$, that is the $\hat\mu_a$
don't scale and the DFs become scale invariant in the considered approximation (with similar conclusions being
previously made, e. g., in \cite{Weinberg:1971mx}). The shear viscosity source term is much simpler.

\section{ The transport coefficients }

After substitution of the $f_k^{(1)}$ with the $\varphi_k$
(\ref{varphi}) into the (\ref{enmomten}) and comparison with the
(\ref{T1}) one finds the formula for the bulk viscosity (cf. \cite{groot}, Chap. VI, Sec. 1)
 \eq{\label{bulkvisc}
 \xi=-\frac13\frac{T}{\sigma(T)}\sum_k x_k(\Delta^{\mu\nu}\pi_{\mu k}\pi_{\nu k},A_k)_k
 =\frac{T}{\sigma(T)}\sum_k x_k(\hat Q_k,A_k)_k,
 }
and for the shear viscosity
 \eq{\label{shearvisc}
 \eta=\frac1{10}\frac{T}{\sigma(T)}\sum_k x_k(\overset{\circ}{\overline{\pi^\mu_k \pi^\nu_k}},
 C_k \overset{\circ}{\overline{\pi_{k\mu} \pi_{k\nu}}})_k,
 }
where the relation ${\Delta^{\mu\nu}_{ ~~\sigma\tau } \Delta_\mu^\sigma \Delta_\nu^\tau = 5}$ is used (cf. \cite{landau10}, Sec. 8). In the
(\ref{bulkvisc}) matching conditions are used, see just below.

In kinetics the conditions that the nonequilibrium perturbations of the DFs do not contribute to the net charge
and the energy-momentum densities are used as a convenient choice and are called the matching conditions, implying also some choice of the flow velocity. They
can reproduce the Landau-Lifshitz condition (\ref{lLcond}) \cite{groot} (Chap. V, Sec. 1). The matching conditions
for the net charge densities can be written as
 \eq{\label{cofchf}
 \sum_k q_{ak} \int \frac{d^3p_k}{(2\pi)^3p_k^0}p_k^\mu U_\mu
 f_k^{(0)}\varphi_k=0,
 }
and for the energy-momentum density can be written as
 \eq{\label{cofemt}
 \sum_k \int \frac{d^3p_k}{(2\pi)^3p_k^0}p_k^\mu p_k^\nu U_\nu f_k^{(0)}\varphi_k=0.
 }
For the special tensorial functions $C_k \overset{\circ}{
\overline{\pi_{k\mu} \pi_{k\nu}}}$ in the (\ref{varphi}) they are
satisfied automatically and for the scalar functions $A_k$ they
can be rewritten in the form (the 3-vector part of the
(\ref{cofemt}) is automatically satisfied)
 \eq{\label{condfit}
 \sum_k q_{ak}x_k (\tau_k, A_k)_k=0, \quad \sum_k x_k (\tau_k^2, A_k)_k=0.
 }
The conditions (\ref{cofchf}) and (\ref{cofemt}) exclude the nonphysical solutions\footnote{Which cannot be
solutions in inhomogeneous systems and are produced just due to shifts in the $T$, $\mu_a$ \cite{landau10} (Sec.
6).} $A_k^{n.ph.} = \sum_a C_a q_{ak} + C \tau_k$ ($C_a$ and $C$ are some constants) of the linearized equations
(\ref{xieqn}) (the (\ref{etaeqn}) don't have nonphysical solutions). One can show explicitly essential
positiveness of the $\xi$ (with help of these matching conditions) and the $\eta$ (within the underlying
assumptions of the BEs), see \cite{groot} (Chap. VI, Sec. 1).

We consider variational (or Ritz) method \cite{groot} (Chap. VI, Sec. 3) allowing to find an approximate solution
of the integral equations (\ref{xieqn}) and (\ref{etaeqn}) in the form of a linear combination of test-functions.
The coefficients next to the test-functions are found from the condition to deliver extremum to some functional,
the first variation of which can reproduce the equations (\ref{xieqn}) and (\ref{etaeqn}). One could take this
functional in the form of some special norm, as in \cite{groot}. Or one can take somewhat different functional,
like in \cite{Arnold:2003zc}, which is a little more convenient, and get the same result. The approximate values
of the viscosities are smaller than the precise ones, being hinted by the applicability of the variational
methods \cite{landau10} (Sec. 10), \cite{Arnold:2003zc}. Questions concerning the uniqueness and existence of the
solution and the convergence of the approximate solution to the precise one are covered in \cite{groot} (Chap.
IX, Secs. 1-2).

The approximate solution of the system of the integral equations (\ref{xieqn}) and (\ref{etaeqn}) are sought in
the form
 \eq{\label{Atestf}
 A_{k}=\sum_{r=0}^{n_1} A_k^r\tau^r_k,
 }
 \eq{\label{Ctestf}
 C_{k}=\sum_{r=0}^{n_2} C_k^r\tau^r_k,
 }
where $n_1$ and $n_2$ set the number of the used test-functions. Test-functions used in \cite{Arnold:2003zc}
would cause less significant digit cancellation in numerical calculations, but there is a need to reduce the
dimension of the 12-dimensional integrals from these test-functions as more as possible to perform the
calculations in a reasonable time. The test-functions in the form of just powers of the $\tau_k$ seem to be the
most convenient for this purpose. As long as particles of the same particle species and different spin states are
undistinguishable, their functions $\varphi_k$ (\ref{varphi}) are equal, and the variational problem is reduced
to the variation of the coefficients $A_{k'}^r$ and $C_{k'}^r$, and the bulk (\ref{bulkvisc}) and the shear
(\ref{shearvisc}) viscosities can be rewritten as
 \eq{\label{finxi}
 \xi=\frac{T}{\sigma(T)}\sum_{k'=1}^{N'}\sum_{r=0}^{n_1}x_{k'}\alpha_{k'}^r A_{k'}^r,
 }
 \eq{\label{fineta}
 \eta=\frac1{10}\frac{T}{\sigma(T)}\sum_{k'=1}^{N'}\sum_{r=0}^{n_2}x_{k'}\gamma_{k'}^rC_{k'}^r.
 }
Applying the variational method one gets the following matrix equations (with the multi-indexes ${(l',s)}$ and
${(k',r)}$) for the bulk and the shear viscosities correspondingly\footnote{One can first derive the same
equations for the $A_k$ and $C_k$, treating them as different functions for all $k$, with the coefficients
$A_{kl}^{rs}$ and $C_{kl}^{rs}$ having the same form as the $A_{k'l'}^{rs}$ and $C_{k'l'}^{rs}$. Then, after
summation of the equations over the spin states of identical particles, and taking ${A_k=A_{k'}}$ and
${C_k=C_{k'}}$, one reproduces the system of equations for the $A_{k'}$ and $C_{k'}$.} \cite{groot} (Chap. VI,
Sec. 3 and Chap. XIII, Sec. 2)
 \eq{\label{ximatreq}
 x_{k'}\alpha_{k'}^r=\sum_{l'=1}^{N'} \sum_{s=0}^{n_1} A_{l'k'}^{sr}A_{l'}^s,
 }
 \eq{\label{etamatreq}
 x_{k'}\gamma_{k'}^r=\sum_{l'=1}^{N'} \sum_{s=0}^{n_2} C_{l'k'}^{sr}C_{l'}^s,
 }
where the introduced coefficients $A_{k'l'}^{rs}$ and
$C_{k'l'}^{rs}$ are
 \eq{\label{A4ind}
 A_{k'l'}^{rs}=x_{k'}x_{l'}[\tau^r,\tau_1^s]_{k'l'}+\delta_{k'l'}x_{k'}\sum_{m'=1}^{N'}
 x_{m'}[\tau^r,\tau^s]_{k'm'},
 }
 \eq{\label{C4ind}
 C_{k'l'}^{rs}=x_{k'}x_{l'}[\tau^r\overset{\circ}{\overline{\pi^{\mu} \pi^{\nu}}},
 \tau_1^s\overset{\circ}{\overline{\pi_{1\mu}\pi_{1\nu}}}]_{k'l'}+\delta_{k'l'}x_{k'}\sum_{m'=1}^{N'}
 x_{m'}[\tau^r\overset{\circ}{\overline{\pi^{\mu} \pi^{\nu}}},
 \tau^s\overset{\circ}{\overline{\pi_{\mu} \pi_{\nu}}}]_{k'm'}.
 }
They are expressed through the collision brackets
\begin{eqnarray}\label{br1}
 [F,G_1]_{kl} &\equiv& \frac{\gamma_{kl}}{T^6(4\pi)^2z_k^2z_l^2K_2(z_k)K_2(z_l)\sigma(T)} \\
 \nonumber &\times& \int_{p_k,p_{1l},{p'}_k,{p'}_{1l}}e^{-\tau_k-\tau_{1l}}(F_k-{F'}_k)G_{1l}W_{kl}.
\end{eqnarray}
The collision brackets $[F,G]_{kl}$ are obtained from the last formula by the replacement of the $G_{1l}$ on the
$G_k$. Due to the time reversibility property of the $W_{kl}$ one can replace the $G_{1l}$ on the
${\frac12(G_{1l}-{G'}_{1l})}$ in the (\ref{br1}). Note that
 \eq{\label{br2pos}
 [\tau^r,\tau^s]_{kl}>0.
 }
It's easy to notice the following symmetries
 \eq{
 [F,G_1]_{kl}=[G,F_1]_{lk}, \quad [F,G]_{kl}=[G,F]_{kl}.
 }
They result in the following symmetric properties: ${A_{k'l'}^{rs} = A_{l'k'}^{sr}}$, ${C_{k'l'}^{rs} =
C_{l'k'}^{sr}}$. Also, the microscopical particle number and energy conservation laws imply for the
$A_{l'k'}^{sr}$ \cite{groot} (Chap. VI, Sec. 3):
 \eq{\label{A4chcons}
 A_{k'l'}^{0s}=0,
 }
 \eq{\label{A4encons}
 \sum_{k'=1}^{N'}A_{k'l'}^{1s}=0.
 }
The (\ref{A4chcons}) together with the $\alpha_k^0=0$ (\ref{elal0}) means that the equations with ${r=0}$ in the
(\ref{ximatreq}) are excluded. From the (\ref{A4encons}) and (\ref{lhsencons}) it follows that each one equation
with ${r=1}$ in the (\ref{ximatreq}) can be expressed through the sum of the other ones, reducing the rank of the
matrix on 1. To solve the matrix equation (\ref{ximatreq}) one eliminates one equation, for example with
${k'=1}$, ${r=1}$. One of coefficients of $A_{l'}^1$ is independent; for example, let it be $A_{1'}^1$. Using the
(\ref{A4encons}), we rewrite the matrix equation (\ref{ximatreq}) in a reduced form as
 \eq{\label{ximatreq2}
 x_{k'}\alpha_{k'}^r=\sum_{l'=2}^{N'}A_{l'k'}^{1r}(A_{l'}^1-A_{1'}^1)
 +\sum_{l'=1}^{N'} \sum_{s=2}^{n_1}A_{l'k'}^{sr}A_{l'}^s.
 }
Then, using the (\ref{elal0}) and the (\ref{lhsencons}), we present the bulk viscosity (\ref{finxi}) in the form
 \eq{\label{finxi2}
 \xi=\frac{T}{\sigma(T)}\sum_{k'=2}^{N'}x_{k'}\alpha_{k'}^1 (A_{k'}^1-A_{1'}^1)
 +\frac{T}{\sigma(T)}\sum_{k'=1}^{N'}\sum_{r=2}^{n_1}x_{k'}\alpha_{k'}^r A_{k'}^r.
 }
Then, the coefficient $A_{1'}^1$ can be eliminated by shifting of other $A_{l'}^1$ and be implicitly used to
satisfy one energy conservation matching condition. The particle number conservation matching conditions are
implicitly satisfied by means of the coefficients $A_{k'}^0$. The first term in the (\ref{finxi2}) is present
only in mixtures. That's why it is small in gases with close to each other masses (and the considered framework
of the BEs) of particles of different species. In gases with very different masses contribution of the first term
in the (\ref{finxi2}) can become dominant.

\section{ Results \label{ResSec}}
\subsection{ The single-component gas \label{singcomsec}}
In the single-component gas, using one test-function, the matrix
equations can be easily solved, and the shear (\ref{fineta}) and
the bulk (\ref{finxi2}) viscosities become (indexes "1" of the
particle species are omitted)
 \eq{\label{etasc}
 \eta=\frac1{10}\frac{T}{\sigma(T)}\frac{(\gamma^0)^2}{C^{00}},
 }
 \eq{\label{xisc}
 \xi=\frac{T}{\sigma(T)}\frac{(\alpha^2)^2}{A^{22}}.
 }
In this approximation the explicit closed-form relativistic formulas for the bulk and the shear viscosities were
obtained in the \cite{anderson}. There the parameter ${a=2r}$. In \cite{groot} (Chap. XI, Sec. 1) they are
written through the parameter ${\sigma=2r^2}$.\footnote{It is the differential cross section for identical
particles. The total cross section is $\sigma_{tot}=\int \frac{d\Omega}2 2r^2=4\pi r^2$.} The results are
 \eq{\label{etaincor}
 \eta=\frac{15}{64\pi}\frac{T}{r^2}\frac{z^2K_2^2(z)\hat h^2}
 {(5z^2+2)K_2(2z)+(3z^3+49z)K_3(2z)},
 }
 \eq{\label{xi}
 \xi=\frac1{64\pi}\frac{T}{r^2}\frac{z^2K_2^2(z)[(5-3\gamma)\hat h-3\gamma]^2}
 {2K_2(2z)+zK_3(2z)},
 }
where ${\gamma=\frac1{c_\upsilon}+1=\frac{z^2+5\hat h-\hat
h^2}{z^2+5\hat h-\hat h^2-1}}$. Though the correct result for the
shear viscosity is
 \eq{\label{eta}
 \eta=\frac{15}{64\pi}\frac{T}{r^2}\frac{z^2K_2^2(z)\hat h^2}
 {(15z^2+2)K_2(2z)+(3z^3+49z)K_3(2z)}.
 }
This result is numerically in agreement with the result in \cite{leeuwen}. To get the (\ref{eta}) and the
(\ref{xi}) the collision brackets in the $C^{00}$ (\ref{C4ind}) and the $A^{22}$ (\ref{A4ind}) can be taken from
\ref{appJ} with ${z_k=z_l=z}$ and the $\gamma^0$ and $\alpha^2$ are defined in the (\ref{algama}). The
discrepancies in the nonrelativistic\footnote{This reproduces the result of Chapman and Enskog in the
nonrelativistic theory for the shear viscosity. The vanishing value of the bulk viscosity is obtained in the
limit $m\rightarrow \infty$ \cite{landau10} (Secs. 8, 10). The result of the vanishing bulk viscosity of a
monoatomic classical gas in the nonrelativistic theory is attributed to James Clerk Maxwell, see
\cite{Weinberg:1971mx}.} ($z\gg 1$) and the ultrarelativistic\footnote{The vanishing value of the bulk viscosity
of a monoatomic classical gas in the ultrarelativistic limit is attributed to I. M. Khalatnikov, see
\cite{landau10} (Sec. 8).} ($z\ll 1$) expansions between the (\ref{etaincor}) and the (\ref{eta}) appear starting
from the first correction. Though the expansions we obtain and the ones in the \cite{anderson} or the
\cite{groot} (Chap. XI, Sec. 1) coincide because they were previously obtained in some other calculations.

The perturbation of the DF $\varphi$ (\ref{varphi}) can be easily found too. We don't know whether this was done
previously, but one may be interested in this result, so we present it too:
 \eq{
 \varphi=\frac1{n\sigma(T)}\left(-(A^0+A^1\tau+A^2\tau^2)\nabla_\mu U^\mu+C^0
 \overset{\circ}{\overline{\pi^\mu \pi^\nu}}
 \overset{\circ}{\overline{\nabla_\mu U_\nu}}\right),
 }
where the $C^0$ is equal to
 \eq{
 C^0=\frac{15}{64\pi}\frac{\sigma(T)}{r^2}\frac{z^2K_2^2(z)\hat h}
 {(15z^2+2)K_2(2z)+(3z^3+49z)K_3(2z)},
 }
and the $A^2$ is equal to
 \eq{
 A^2=\frac1{64\pi}\frac{\sigma(T)}{r^2}\frac{z^2K_2^2(z)[(5-3\gamma)\hat h-3\gamma]}
 {2K_2(2z)+zK_3(2z)}.
 }
The $A^0$ and the $A^1$ are used to satisfy the matching
conditions (\ref{condfit}) and are equal to
 \eq{
 A^0=A^2\frac{a^2a^4-(a^3)^2}{\Delta_A}, \quad
 A^1=A^2\frac{a^2a^3-a^1a^4}{\Delta_A}, \quad
 \Delta_A\equiv a^1 a^3-(a^2)^2,
 }
where the $a^r$ are defined in the (\ref{algama}). In the nonrelativistic limit (${z\gg 1}$) one has
 \eq{
 \varphi=\frac{5\pi e^{z-\hat\mu}}{32\sqrt2 T^3 z^2 r^2}
 \left(-(\tau^2+2z\tau-z^2)\nabla_\mu U^\mu+
 2\overset{\circ}{\overline{\pi^\mu \pi^\nu}}
 \overset{\circ}{\overline{\nabla_\mu U_\nu}}\right).
 }
In the ultrarelativistic limit ($z\ll 1$) one has
 \eq{
 \varphi=\frac{\pi e^{-\hat\mu}}{480 T^3 r^2}\left(-5z^2(\tau^2+8\tau-12)\nabla_\mu
 U^\mu+36\overset{\circ}{\overline{\pi^\mu \pi^\nu}}
 \overset{\circ}{\overline{\nabla_\mu U_\nu}}\right).
 }
Note that, although the shear viscosity diverges for ${T\rightarrow \infty}$, the perturbative expansion over the
gradients does not break down because the $\varphi$ does not diverge (it tends to zero, conversely).

The phenomenological formula, coming from the momentum transfer considerations in the kinetic-molecular theory,
for the shear viscosity is ${\eta_{ph}\propto l n \langle \abs{\vec p} \rangle}$ (with the coefficient of
proportionality $\sim 1$), where $\langle \abs{\vec p} \rangle$ is the average relativistic momentum, $l$ is the
mean free path. It gives the correct leading $m$ and $T$ parameter dependence of the (\ref{eta}) with quite a
precise coefficient\footnote{This formula is justified only for rarefied systems where the IG equation of state
is applicable.}. The mean free path can be estimated as ${l\sim 1/(\sigma_{tot}n)}$ (see Sec. \ref{appmfp}).
Choosing the coefficient of proportionality to match the nonrelativistic limit one gets \cite{Gorenstein:2007mw}
 \eq{\label{etaph}
 \eta_{ph}=\frac{5}{64\sqrt{\pi}}\frac{\sqrt{mT}}{r^2}\frac{K_{5/2}(m/T)}{K_2(m/T)}.
 }
If the bulk viscosity is expressed as ${\xi_{ph}\propto l n \langle \abs{\vec p} \rangle}$, the coefficient of
proportionality is not $\sim 1$. In the nonrelativistic limit it is $25/(512 \sqrt2 z^2)$ and in the
ultrarelativistic limit it is $z^4/(864 \pi)$. To reproduce these asymptotical dependencies the bulk viscosity
should be proportional to the second power of the averaged product of the source term $\hat Q$ and the $\tau$,
that is to the $(\alpha^2)^2$.

If a system has no charges, then terms proportional to the $\tau_k$ in the (\ref{Qsource}) are absent, and the
$R$ quantity gets another form. This results in quite different values of the $\alpha_k^r$. In particular, for
the single-component gas in the case ${z\gg 1}$ one gets
 \eq{\label{alfrac1}
 \frac{(\alpha^2)^2|_{q_{11}=0}}{(\alpha^2)^2|_{q_{11}=1}}=\frac{4 z^4}{25}+...,
 }
and in the case ${z\ll 1}$ one gets
 \eq{\label{alfrac2}
 \frac{(\alpha^2)^2|_{q_{11}=0}}{(\alpha^2)^2|_{q_{11}=1}}=4+... .
 }
In both cases these estimates suppose enhancement of the bulk viscosity (\ref{xisc}) if the number-changing
processes are not negligible.

\subsection{ The binary mixture \label{binmixsec}}

The mixture of two species with masses $m_1$, $m_2$ and the different classical elastic differential constant
cross sections $\sigma^{cl}_{11}$, ${\sigma^{cl}_{12} = \sigma^{cl}_{21}}$, $\sigma^{cl}_{22}$ is considered in
this section (only formal expressions can be found in the \cite{groot}, Chap. VI, Sec. 3). Using the
(\ref{fineta}) with ${n_2=0}$ and solving the matrix equation (\ref{etamatreq}) one gets for the shear viscosity
 \eq{
 \eta=\frac{T}{10\sigma(T)}\frac1{\Delta_\eta}[(x_{1'}\gamma_1^0)^2C_{2'2'}^{00}-
 2x_{1'}x_{2'}\gamma_1^0\gamma_2^0C_{1'2'}^{00}+(x_{2'}\gamma_2^0)^2C_{1'1'}^{00}],
 }
where ${\Delta_\eta = C_{1'1'}^{00}C_{2'2'}^{00} - (C_{1'2'}^{00})^2}$. The collision brackets for the
$C_{k'l'}^{00}$ (\ref{C4ind}) can be found in \ref{appJ} and the $\gamma_k^0$ are defined in the (\ref{algama}).

In the important limiting case when one mass is large ${z_2\gg 1}$ ($g_2$ and $\hat \mu_2$ are finite so that
${x_{2'}\ll 1}$) and another mass is finite one can perform asymptotic expansion of the special functions. Then,
one has ${x_{1'} \propto O(1)}$, ${\gamma_1^0 \propto O(1)}$, ${x_{2'} \propto O(e^{-z_2}z_2^{3/2})}$,
${\gamma_2^0 \propto O(z_2)}$. The collisions of light and heavy particles dominate over the collisions of heavy
and heavy particles in the $C_{2'2'}^{00}$, and one has ${[\overset{\circ}{ \overline{\pi^{\mu} \pi^{\nu}} },
\overset{\circ}{ \overline{\pi_{\mu} \pi_{\nu}} }]_{21} \propto O(z_2)}$, ${C_{2'2'}^{00} \propto
O(e^{-z_2}z_2^{5/2})}$. In the $C_{1'1'}^{00}$ the collisions of light and light particles dominate, and one gets
${C_{1'1'}^{00} \propto O(1)}$. And $[\overset{\circ}{ \overline{\pi^{\mu} \pi^{\nu}} }, \overset{\circ}{
\overline{\pi_{1\mu} \pi_{1\nu}} }]_{12} \propto O(1)$, ${C_{1'2'}^{00} \propto O(e^{-z_2}z_2^{3/2})}$. In the
shear viscosity the first nonvanishing contribution is the single-component shear viscosity (\ref{eta}), where
one should take ${r^2=\sigma^{cl}_{11}}$ and ${z=z_1}$. The next correction is
 \eq{
 \Delta \eta=z_2^{5/2}e^{-z_2}\frac{3 Tg_2 e^{z_1-\hat\mu_1+\hat\mu_2}}
 {64\sqrt{2\pi}(3+3z_1+z_1^2)g_1\sigma^{cl}_{12}}.
 }
The approximate formula \cite{Gorenstein:2007mw}
 \eq{
 \eta=\sum_k\eta_kx_k,
 }
where $\eta_k$ is given by the (\ref{eta}) or the (\ref{etaph})
with mass $m_k$ and cross section $\sigma^{cl}_{kk}$, would give
somewhat different heavy mass power dependence $O(e^{-z_2}z_2^2)$.

Using the (\ref{finxi2}) with ${n_1=1}$ and solving the matrix
equation (\ref{ximatreq2}) one gets for the bulk viscosity
 \eq{
 \xi=\frac{T}{\sigma(T)}\frac{(x_{2'}\alpha_2^1)^2}{A_{2'2'}^{11}}=
 \frac{T}{\sigma(T)}\frac{x_{1'}x_{2'}\alpha_1^1\alpha_2^1}{A_{1'2'}^{11}}.
 }
Using the definition of the $A_{2'2'}^{11}$ (\ref{A4ind}) and the fact ${[\tau,\tau_1]_{kl} + [\tau,\tau]_{kl} =
0}$ (\ref{br211}) one gets ${A_{2'2'}^{11} = x_{1'}x_{2'}[\tau,\tau]_{12}}$. Using the (\ref{br2pos}) one gets
${[\tau,\tau]_{12} > 0}$. Then, using ${x_{1'}\alpha_1^1 + x_{2'}\alpha_2^1 = 0}$, coming from the
(\ref{lhsencons}), the bulk viscosity can be rewritten as
 \eq{
 \xi=\frac{T}{\sigma(T)}\frac{x_{2'}(\alpha_2^1)^2}{x_{1'}[\tau,\tau]_{12}}
 =\frac{T}{\sigma(T)}\frac{x_{1'}(\alpha_1^1)^2}{x_{2'}[\tau,\tau]_{12}}>0.
 }
The collision bracket $[\tau,\tau]_{12}$ can be found in \ref{appJ}, and the $\alpha_k^1$ are defined in the
(\ref{algama}).

In the limiting case ${z_2\gg 1}$ one has ${x_{1'} \propto O(1)}$, ${x_{2'} \propto O(e^{-z_2}z_2^{3/2})}$,
$\alpha_1^1 \propto O(e^{-z_2}z_2^{3/2})$, ${\alpha_2^1 \propto O(1)}$, ${A_{22}^{11} \propto A_{12}^{11} \propto
O(e^{-z_2}z_2^{1/2})}$, ${[\tau,\tau]_{12} \propto O(z_2^{-1})}$. Then, for the bulk viscosity one gets
 \eq{
 \xi=e^{-z_2}z_2^{5/2}\frac{g_2 T e^{-\hat\mu_1+\hat\mu_2+z_1}[2 z_1^2-5-2\hat h_1^2
 +10 \hat h_1]^2}{128 \sqrt{2 \pi } g_1 \sigma^{cl}_{12}(z_1^2+3 z_1+3)[z_1^2-1-\hat h_1^2+5 \hat h_1]^2}+....
 }

\subsection{ The collision rates and the mean free paths \label{appmfp}}

The quantity $\frac{W_{k'l'}} {p_{k'}^0 p_{1l'}^0 {p'}_{k'}^0 {p'}_{1l'}^0} d^3{p'}_{k'} d^3{p'}_{1l'}$, which
enters the elastic collision integral (\ref{ckelgroot}), represents the probability of scattering per unit time
times unit volume for two particles which had momentums $\vec p_{k'}$ and $\vec p_{1l'}$ before scattering and
momentums in the ranges ${(\vec {p'}_{k'}, \vec {p'}_{k'} + d\vec {p'}_{k'})}$ and ${(\vec {p'}_{1l'}, \vec
{p'}_{1l'} + d\vec {p'}_{1l'})}$ after the scattering. The quantity $ g_{k'} \frac{d^3p_{k'}}{ (2\pi)^3 } f_{k'}$
represents the number of particles per unit volume, which have momentums in the range ${(\vec p_{k'}, \vec p_{k'}
+ d\vec p_{k'})}$. The number of collisions of particles of the $k'$-th species with particles of the $l'$-th
species per unit time per unit volume is then\footnote{It represents some sum over all possible collisions. In
the case of the same species one factor $\gamma_{k'l'}$ just cancels the double counting in momentum states after
scattering and another factor $\gamma_{k'l'}$ also reflects the fact that scattering takes place for
${n_{k'}\choose 2} \approx \frac12n_{k'}^2$ pairs of undistinguishable particles in a given unit volume.}
 \eq{\label{totratekl}
 \widetilde R^{el}_{k'l'}\equiv g_{k'}g_{l'}\frac{\gamma_{k'l'}^2}{(2\pi)^6}\int
 \frac{d^3p_{k'}}{p_{k'}^0}\frac{d^3p_{1l'}}{p_{1l'}^0}\frac{d^3p'_{k'}}{{p'}_{k'}^0}
 \frac{d^3p'_{1l'}}{{p'}_{1l'}^0}f_{k'}^{(0)}f_{1l'}^{(0)}W_{k'l'}.
 }
To get the corresponding number of collisions of particles of the
$k'$-th species with particles of the $l'$-th species per unit
time \emph{per particle of the $k'$-th species}, $R^{el}_{k'l'}$,
one has to divide the (\ref{totratekl}) on the
$\gamma_{k'l'}n_{k'}$ (recall that ${n_{k'} \propto g_{k'}}$ by
definition), which is the number of particles of the $k'$-th
species per unit volume divided on the number of particles of the
$k'$-th species taking part in the given type of reaction (2 for
binary elastic collisions, if particles are identical, and 1
otherwise). This rate can be directly obtained averaging the
collision rate with fixed momentum $p_k$ of the $k$-th particle
species
 \eq{
 \mc R^{el}_{kl'}\equiv g_{l'}\gamma_{kl'}\int
 \frac{d^3p_{1l'}}{(2\pi)^3}d^3p'_{k}d^3p'_{1l'}f_{1l'}^{(0)}
 \frac{W_{kl'}}{p_{k}^0p_{1l'}^0{p'}_{k}^0{p'}_{1l'}^0},
 }
over the momentum with the probability distribution $\frac{d^3p_{k}} {(2\pi)^3} \frac{f_{k}}{n_k}$ (and spin
states, which is trivial):
 \eq{
 R^{el}_{k'l'}\equiv g_{k'}g_{l'}\frac{\gamma_{k'l'}}{(2\pi)^6n_{k'}}\int
 \frac{d^3p_{k'}}{p_{k'}^0} \frac{d^3p_{1l'}}{p_{1l'}^0}
 \frac{d^3p'_{k'}}{{p'}_{k'}^0}\frac{d^3p'_{1l'}}{{p'}_{1l'}^0}f_{k'}^{(0)}
 f_{1l'}^{(0)}W_{k'l'}=\frac{\widetilde R^{el}_{k'l'}}{\gamma_{k'l'}n_{k'}}.
 }
So that to get the mean rate of the elastic collisions per
particle of the $k'$-th species with all particles in the system
one can just integrate the sum of the gain terms in the collision
integral (\ref{ckelgroot}) over $\frac{d^3p_k}{(2\pi)^3p_k^0n_k}$
and average it over spin:
 \eq{
 R_{k'}^{el}\equiv \sum_{l'} R_{k'l'}^{el}.
 }
One can express the $\widetilde R_{k'l'}^{el}$ through the $J_{kl}^{(0, 0, 0, 0, 0| 0, 0)}$ integrals from
\ref{appJ} as
 \eq{\label{Rklel}
 \widetilde R_{k'l'}^{el}=\gamma_{k'l'}\sigma(T)n_{k'} n_{l'}
 J_{k'l'}^{(0, 0, 0, 0, 0| 0, 0)}.
 }
Then, the (\ref{Rklel}) becomes
\begin{eqnarray}
 \widetilde R_{k'l'}^{el} &=& g_{k'}g_{l'}\gamma_{k'l'}\frac{2\sigma^{cl}_{k'l'} T^6}{\pi^3}
  [(z_{k'}-z_{l'})^2 K_2(z_{k'}+z_{l'}) \\
 \nonumber &+& z_{k'} z_{l'}(z_{k'}+z_{l'})K_3(z_{k'}+z_{l'})],
\end{eqnarray}
where $\sigma^{cl}_{k'l'}$ is the classical elastic differential constant cross section of scattering of a
particle of the $k'$-th species on particles of the $l'$-th species. For the case of large temperature or when
both masses are small, ${z_{k'} \ll 1}$ and ${z_{l'} \ll 1}$, one has expansion
 \eq{
 \widetilde R_{k'l'}^{el}=g_{k'}g_{l'}\gamma_{k'l'}
 \frac{4 \sigma^{cl}_{k'l'} T^6}{\pi^3}
 \left(1-\frac14(z_{k'}^2+z_{l'}^2)+...\right).
 }
For the case of small temperature or when both masses are large,
${z_{k'}\gg 1}$ and ${z_{l'}\gg 1}$, one has expansion
\begin{eqnarray}
 \widetilde R_{k'l'}^{el} &=& g_{k'}g_{l'}\gamma_{k'l'}\frac{\sqrt{2}\sigma^{cl}_{k'l'} T^6 z_{k'} z_{l'}
 \sqrt{z_{k'}+z_{l'}} e^{-z_{k'}-z_{l'}}}{\pi^{5/2}} \\
 \nonumber &\times& \left(1+\frac{8 z_{k'}^2+19 z_{k'} z_{l'}+8 z_{l'}^2}{8 z_{k'} z_{l'} (z_{k'}+z_{l'})}+...\right).
\end{eqnarray}
For the case when only one mass is large, ${z_{l'}\gg 1}$, one has somewhat different expansion
\begin{eqnarray}
 \widetilde R_{k'l'}^{el} &=& g_{k'}g_{l'}\gamma_{k'l'}\frac{\sqrt{2}
 \sigma^{cl}_{k'l'} T^6 (z_{k'}+1) z_{l'}^{3/2} e^{-z_{k'}-z_{l'}}}{\pi^{5/2}} \\
 \nonumber &\times& \left(1+\frac{4 z_{k'}^2+15 z_{k'}+15}{8 z_{k'}+8}z_{l'}^{-1}...\right).
\end{eqnarray}
The $\sigma(T) J_{k'l'}^{(0, 0, 0, 0, 0| 0, 0)}$ in the (\ref{Rklel}) can be replaced in the ultrarelativistic
limit with $4\pi \sigma^{cl}_{k'l'} \langle \abs{\vec v_{k'}}\rangle$ and in the nonrelativistic limit with $4\pi
\sigma^{cl}_{k'l'} \langle \abs{\vec v_{rel,k'l'}}\rangle = 4\pi \sigma^{cl}_{k'l'} \langle \abs{\vec v_{k'}}
\rangle \sqrt{1+m_{k'}/m_{l'}}$, where $\langle \abs{\vec v_{k'}}\rangle$ is the mean modulus of particle's
velocity of the $k'$-th species,
 \eq{\label{avvel}
 \langle \abs{\vec v_{k'}}\rangle=\frac{\int d^3p_{k'} \frac{|\vec p_{k'}|}{p_{k'}^0}
 f^{(0)}_{k'}(p_{k'})}{\int d^3p_{k'} f^{(0)}_{k'}(p_{k'})}
 =\frac{2 e^{-z_{k'}}(1+z_{k'})}{z_{k'}^2K_2(z_{k'})}
 =\sqrt{\frac{8}{\pi z_{k'}}}\frac{K_{3/2}(z_{k'})}{K_{2}(z_{k'})},
 }
and $\langle \abs{\vec v_{rel,k'l'}}\rangle$ is the mean modulus of the relative velocity, which coincides with
the $\langle \abs{\vec v_{k'}}\rangle$ at high temperatures. Then, the resultant collision rate $R^{el}_{k'}$
would reproduce simple nonrelativistic collision rates know in the kinetic-molecular theory \cite{boltzmann}
(Sec. 9). To get a (approximate) mean free time one has just to invert the $R^{el}_{k'}$,
$t^{el}_{k'}=1/R^{el}_{k'}$. A (approximate) mean free path $l^{el}_{k'}$ can be obtained after multiplication of
it on the $\langle \abs{\vec v_{k'}}\rangle$:
 \eq{\label{lel}
 l^{el}_{k'}=\frac{\langle \abs{\vec v_{k'}}\rangle}{R^{el}_{k'}}.
 }
For the single-component gas one gets
 \eq{\label{mfpsc}
 l^{el}_{1'}=\frac{\langle \abs{\vec v_{1'}}\rangle}{R^{el}_{1'1'}}=
 \frac{\pi e^{-z_{1}} (z_{1}+1)}{g_{1}4 \sigma^{cl}_{11} T^3 z_{1}^3 K_3(2 z_{1})}.
 }
The nonrelativistic limit of the (\ref{mfpsc}) with the ${g_1=1}$
coincides with the same limit of the formula
 \eq{
 l^{el}_{1}=\frac{\langle \abs{\vec v_1}\rangle}{4\pi\sigma^{cl}_{11}n_1
 \langle \abs{\vec v_{rel}}\rangle}=\frac{1}{4\pi\sigma^{cl}_{11}n_1\sqrt2},
 }
which is the mean free path formula coming from the nonrelativistic kinetic-molecular theory, obtained by Maxwell
\cite{boltzmann} (Secs. 10, 9). The ultrarelativistic limit of the (\ref{mfpsc}) with the ${g_1=1}$ coincides
with the same limit of the formula
 \eq{
 l_1^{el}=\frac{1}{4\pi\sigma^{cl}_{11}n_1}.
 }

\section{Discussion \label{DiscSec}}

The presented formulas can be considered as quite precise ones for rarefied gases with short-range interactions.
However, for the bulk viscosity one should be sure that the approximation of only the elastic processes is a good
one. Our estimations indicate the considerable increase of the bulk viscosity if inelastic processes are
introduced. With the choice of the chemical potentials and the temperature as independent thermodynamic variables
we find the bulk viscosity source term (\ref{Qsource}) to have a more convenient and universal form. We believe
this piece of methodology is a new result. There are analogical expressions for multicomponent gases, e. g., as
in \cite{Khvorostukhin:2010aj}, but our expression does not require further transformations if the
thermodynamic functions are known as functions of the temperature and the chemical potentials (which is usually so).
Also any sort of analysis or obtaining of a partial case, like the one of zero chemical potentials, might be
more easily done having the chemical potentials and the temperature as independent variables. Switching between
the two qualitatively different cases of maximal particle number conservation and nonconservation (and
intermediate ones) is realized through one quantity, the matrix $q_{ak}$.

The approximation of the Maxwell-Boltzmann (classical) statistics and only the elastic processes allows one to
obtain relatively simple analytical closed-form expressions. We present the correct form of the single-component
shear viscosity, though the deviations are small and appear only in the corrections of the nonrelativistic and
the ultrarelativistic expansions. We have used Mathematica package \cite{math}, which allowed to avoid such
omissions in this and more bulky expressions. The single-component and the binary mixture formulas can be used
for estimations or checkups of the viscosities in mixtures, besides their direct applications. We also present
simple expressions for the collision rates and approximate formulas for the mean free paths and times, which one
might need for justification of the applicability of the hydrodynamical description. We believe that all these
formulas (except for the singe-component bulk viscosity) are new too.

We present the lowest orders collision brackets for the viscosities in the mixtures and the $J$-integrals through
which one can obtain the collision brackets of other orders (with ease using an algebraic manipulation package).
To calculate any transport coefficient using a variational method (allowing to control precision) in the lowest
order approximation in a mixture with a very large number of components $N'$ one would need to calculate
$O({N'}^2)$ 12-dimensional $J$-integrals, being the main difficulty, if only the elastic collisions are
considered. Fortunately, it's possible to simplify these integrals considerably, expressing them through the
known special functions. This allows to perform calculations of the viscosities in gases with many particle
species in a reasonable time. We consider this the main result.

If there are nonconstant cross sections, then one can describe the viscosities approximately using averaged
(temperature dependent) cross sections with the formulas mentioned above. We postpone to the future our possible
investigations of the deviations due to application of the averaged cross sections. If for some energy
dependencies of the cross sections the deviations are large, improved definitions of the averaged cross sections
may exist. The next-to-leading order corrections in the form of thermal masses and thermal cross sections, as
well as the beta function contribution to the bulk viscosity source term at high enough temperatures, as in
the \cite{jeon} can be relatively easily taken into account if they are calculated. The main applications of the results
of this paper are to the hydrodynamical description of the hadronic phase created in heavy ion collisions
\cite{Gorenstein:2007mw, Moroz:2011vn, Fogaca:2013cma}, where the particles are relativistic ones and there are
many species of them, or in the same but simplified description \cite{davesne, Dobado:2009ek, dobado}. The
formulas mentioned above may have similar applications to other fields of physics, like cosmology
\cite{Weinberg:1971mx, Sawyer:2006ju, Das:2008mj, BasteroGil:2012zr}, solid state physics \cite{phonon,
graphene}, studies of atmosphere \cite{air}, etc.

\section{Conclusions \label{ConcSec}}

We have obtained formulas for the shear and the bulk viscosities in mixtures with constant cross sections. They
allow to conduct fast computations for mixtures with large number of particle species. These expressions can be
used in many fields of physics. These formulas represent the relativistic generalization of the formulas for the
nonrelativistic hard spheres model. Additionally, we have presented explicit relatively simple expressions (which
is a benefit for applications) for different auxiliary quantities, like the collision rates.

\section{Acknowledgments}
The author is grateful to the referees for comments on the improvement of this paper's form.

\appendix

\section{The calculation of the collision brackets \label{appJ}}
The momentum parametrization and the most of transformations of the 12-dimensional integrals considered below are
taken from the \cite{groot} (Chap. XI and XIII). There the reader can find some assignments appearing below.

There is a need to calculate the following integrals
 \begin{eqnarray}\label{Jint0}
  \nonumber &~& J_{kl}^{(a,b,d,e,f|q,r)} \equiv \frac{\gamma_{kl}}{T^6(4\pi)^2z_k^2z_l^2K_2(z_k)K_2(z_l)
  \sigma(T)} \int_{p_k,p_{1l},{p'}_k,{p'}_{1l}} \\
  \nonumber &~& \times e^{-P\cdot U/T}(1+\alpha_{kl})^q(1-\alpha_{kl})^r \left(\frac{P^2}{T^2}\right)^a
  \left(\frac{P\cdot U}{T}\right)^b\left(\frac{Q\cdot U}{T}\right)^d \\
  &~& \times \left(\frac{Q'\cdot U}{T}\right)^e\left(\frac{-Q\cdot {Q'}}{T^2}\right)^f W_{kl}.
 \end{eqnarray}
Using some  nontrivial transformations, described in more details in \cite{groot} (Chap. XI and XIII), we arrive
at
 \begin{eqnarray}\label{Jint}
  \nonumber &~& J_{kl}^{(a,b,d,e,f|q,r)}=\frac{\pi(d+e+1)!!\sigma^{(d,e,f)}_{1kl}}{z_k^2z_l^2K_2(z_k)K_2(z_l)}
  \sum_{q_1=0}^q\sum_{r_1=0}^r\sum_{k_2=0}^{\frac{d+e}2+f+1}\sum_{k_3=0}^{\frac{d+e}2+f+1}
  \sum_{h=0}^{[b/2]} \\
  \nonumber &~&\times (z_k+z_l)^{2(q_1+r_1+k_2+k_3)}\left(\frac{z_k-z_l}{z_k+z_l}\right)^{q_1+r_1+2k_3}
  (-1)^{r_1+k_2+k_3+h}(2h-1)!! \\
  &~& \times \binom{b}{2h}\binom{q}{q_1}\binom{r}{r_1}
  \binom{\frac{d+e}2+f+1}{k_2}\binom{\frac{d+e}2+f+1}{k_3} \\
  \nonumber &~& \times I\left(2(a+f-q_1-r_1-k_2-k_3)+3,b+\frac{d+e}2-h+1,z_k+z_l\right),
 \end{eqnarray}
where
 \eq{
 \sigma_{1kl}^{(d,e,f)}=\frac{\sigma^{cl}_{kl}}{\sigma(T)}
 \sum_{g=0}^{\min(d,e)} \sigma^{(f,g)} K(d,e,g),
 }
where ${\sigma^{cl}_{kl} = \gamma_{kl}\sigma_{kl}}$ is the
classical elastic differential constant cross section. The
$\sigma^{(f,g)}$ is equal to the real, nonzero and non-diverging
value (for any non-negative integer $g$)
 \eq{\label{sigmafg}
 \sigma^{(f,g)}=\frac{2g+1}{2}\int_{-1}^1 dx x^f P_g(x)=(2g+1) \frac{f!}{(f-g)!!(f+g+1)!!},
 }
if the difference ${f-g}$ is even and ${g \leq f}$. Above the
$P_{\text{g}}(x)$ is the Legendre polynomial. The $K(d,e,g)$ is
equal to the real, the nonzero and non-diverging quantity (for any
non-negative integer $g$)
 \eq{
 K(d,e,g)=\frac{d! e!}{(d-g)!! (d+g+1)!! (e-g)!! (e+g+1)!!},
 }
if ${g \leq \min(d,e)}$ and both the ${d-g}$ and the ${e-g}$ are
even (which also implies that ${d+e}$ is even). The ${[...]}$
denotes the integer part. The integral $I$ is
 \eq{\label{IIntdef}
 I(r,n,x)\equiv x^{r+n+1}\int_1^\infty du u^{r+n}K_n(xu).
 }
Also, there is the following frequently used combination of the $J$ integrals
\begin{eqnarray}
 \nonumber {J'}_{kl}^{(a,b,d,e,f|q,r)} &\equiv& \sum_{u=0}^f (-1)^u\binom{f}{u}
 (2z_k)^{2(f-u)}J_{kl}^{(a+k,b,d+e,0,0|q+2u,r)} \\
 &-& J_{kl}^{(a,b,d,e,f|q,r)}.
\end{eqnarray}
The first term in the difference is obtained by the replacement of the $Q'$ on the $Q$ everywhere except for the
$W_{kl}$. Using this fact, the $J'$ can be rewritten in the form
 \eq{\label{Jpint}
 {J'}_{kl}^{(a,b,d,e,f|q,r)}=\frac{\sigma^{(d,e,f)}_{kl}}{(d+e+1)\sigma^{(d,e,f)}_{1kl}}J_{kl}^{(a,b,d,e,f|q,r)},
 }
where
 \begin{eqnarray}\label{sigmadef}
  \nonumber \sigma^{(d,e,f)}_{kl}&=&\frac{\sigma^{cl}_{kl}}{\sigma(T)}(d+e+1)\left(K(d+e,0,0)\sigma^{(0,0)}
  -\sum_{g=0}^{\min(d,e)}K(d,e,g)\sigma^{(f,g)}\right)\\
  &=&\frac{\sigma^{cl}_{kl}}{\sigma(T)}\left(1-(d+e+1)\sum_{g=0}^{\min(d,e)}K(d,e,g)\sigma^{(f,g)}\right).
 \end{eqnarray}
There is a recurrence relation for the integral $I$
(\ref{IIntdef}) \cite{groot} (Chap. XI, Sec. 1):
\begin{eqnarray}\label{recrel}
 I(r,n,x) &=& (r-1)(r+2n-1)I(r-2,n,x) \\
 \nonumber &+& (r-1)x^{r+n-1}K_n(x)+x^{r+n}K_{n+1}(x).
\end{eqnarray}
For the calculations one needs only the integrals $I(r,n,x)$ with the positive values of the $n$ and the odd
values of the $r$. If ${r \geq -2n+1}$, the $I$ integrals can be expressed through the Bessel functions $K_n(x)$,
using the (\ref{recrel}), when ${r=1}$ or ${r=-2n+1}$. Then, using the recurrence relation for the $K_n(x)$
\cite{luke},
 \eq{\label{Krecrel}
 K_{n+1}(x)=K_{n-1}(x)+\frac{2n}{x}K_n(x),
 }
the final result can be expressed through a couple of Bessel
functions. If ${r\leq-2n-1}$, then the recurrence relation
(\ref{recrel}) becomes singular if one tries to express the
$I(r,n,x)$ through the $I(-2n+1,n,x)$. Using the (\ref{recrel}),
the $I$ integrals with ${r \leq -2n-1}$ can be expressed through
the integrals $G(n,x)$
 \eq{\label{Gdef}
 G(n,x)\equiv I(-2n-1,n,x)=x^{-n}\int_1^\infty du u^{-n-1}K_n(xu).
 }
There is a recurrence relation for the $G(n,x)$:
 \eq{\label{Grecrel}
 G(n,x)=-\frac1{2n}(G(n-1,x)-x^{-n} K_n(x)).
 }
It can be easily proved by the integration by parts of the
(\ref{Gdef}) and using the following relation for the $K_n(x)$
\cite{luke}
 \eq{\label{dKdx}
 \frac{\p }{\p x}K_n(x)=-\frac{n}{x} K_n(x)-K_{n-1}(x).
 }
It is found that collision brackets have the simplest form if they are expressed through $G(n,x)$ with ${n=3}$ or
${n=2}$ and the Bessel functions $K_3(x)$ and $K_2(x)$ or $K_2(x)$ and $K_1(x)$. It was chosen to take ${G(x)
\equiv G(3,x)}$ and $K_3(x)$, $K_2(x)$. The $G(x)$ can be expressed through the Meijer function \cite{meijer}
 \eq{
 G(x)=\frac{1}{32}
 G_{1,3}^{3,0}\left((x/2)^2\left|
 \begin{array}{c}
  1 \\
  -3,0,0
 \end{array}\right.
 \right).
 }
The needed scalar collision brackets can be expressed through the
$J'$ as
 \eq{
 [\tau^r,\tau^s_1]_{kl}=\frac1{2^{r+s}}\sum_{u=1}^r\sum_{\upsilon=1}^s
 (-1)^\upsilon \binom{r}{u} \binom{s}{\upsilon} {J'}_{kl}^{(0,r+s-u-\upsilon,u,
 \upsilon,0|r-u,s-\upsilon)},
 }
 \eq{
 [\tau^r,\tau^s]_{kl}=\frac1{2^{r+s}}\sum_{u=1}^r\sum_{\upsilon=1}^s
 \binom{r}{u} \binom{s}{\upsilon} {J'}_{kl}^{(0,r+s-u-\upsilon,u,
 \upsilon,0|r+s-u-\upsilon,0)},
 }
and the needed tensorial collision brackets can be expressed as
 \begin{eqnarray}
  &~& [\tau^r\overset{\circ}{\overline{\pi^{\mu} \pi^{\nu}}},
  \tau_1^s\overset{\circ}{\overline{\pi_{1\mu} \pi_{1\nu}}}]_{kl}=
  \frac1{2^{r+s+4}}\sum_{n_1=0}^r\sum_{n_2=0}^s \binom{s}{n_2}\binom{r}{n_1}(-1)^{s-n_2} \\
  \nonumber &\times& ({J'}_{kl}^{(2,n_1+n_2,r-n_1,s-n_2,0|2+n_1,2+n_2)}+
  2{J'}_{kl}^{(1,n_1+n_2,r-n_1,s-n_2,1|1+n_1,1+n_2)} \\
  \nonumber &+& {J'}_{kl}^{(0,n_1+n_2,r-n_1,s-n_2,2|n_1,n_2)})
  -\frac1{2^{r+s+3}}\sum_{n_1=0}^{r+1}\sum_{n_2=0}^{s+1} \binom{s+1}{n_2}\binom{r+1}{n_1} \\
  \nonumber &\times& (-1)^{s+1-n_2}({J'}_{kl}^{(1,n_1+n_2,r+1-n_1,s+1-n_2,0|1+n_1,1+n_2)} \\
  \nonumber &+&{J'}_{kl}^{(0,n_1+n_2,r+1-n_1,s+1-n_2,1|n_1,n_2)})+\frac23[\tau^{r+2},\tau^{s+2}_1]_{kl} \\
  \nonumber &+& \frac13 z_l^2 [\tau^{r+2},\tau^{s}_1]_{kl} + \frac13 z_k^2 [\tau^{r},\tau^{s+2}_1]_{kl}- \frac13 z_k^2 z_l^2
  [\tau^{r},\tau^{s}_1]_{kl},
 \end{eqnarray}
 \begin{eqnarray}
  &~& [\tau^r\overset{\circ}{\overline{\pi^{\mu} \pi^{\nu}}},
  \tau^s\overset{\circ}{\overline{\pi_{\mu} \pi_{\nu}}}]_{kl}=
  \frac1{2^{r+s+4}}\sum_{n_1=0}^r\sum_{n_2=0}^s \binom{s}{n_2}\binom{r}{n_1} \\
  \nonumber &\times& ({J'}_{kl}^{(2,n_1+n_2,r-n_1,s-n_2,0|4+n_1+n_2,0)}
  - 2{J'}_{kl}^{(1,n_1+n_2,r-n_1,s-n_2,1|2+n_1+n_2,0)} \\
  \nonumber &+& {J'}_{kl}^{(0,n_1+n_2,r-n_1,s-n_2,2|n_1+n_2,0)})
  - \frac1{2^{r+s+3}}\sum_{n_1=0}^{r+1}\sum_{n_2=0}^{s+1}\binom{s+1}{n_2}\binom{r+1}{n_1} \\
  \nonumber &\times& ({J'}_{kl}^{(1,n_1+n_2,r+1-n_1,s+1-n_2,0|2+n_1+n_2,0)} \\
  \nonumber &-& {J'}_{kl}^{(0,n_1+n_2,r+1-n_1,s+1-n_2,1|n_1+n_2,0)}) + \frac23[\tau^{r+2},\tau^{s+2}]_{kl} \\
  \nonumber &+& \frac13 z_k^2 [\tau^{r+2},\tau^{s}]_{kl} + \frac13 z_k^2 [\tau^{r},\tau^{s+2}]_{kl}
  - \frac13 z_k^4[\tau^{r},\tau^{s}]_{kl}.
 \end{eqnarray}
Below some lowest orders collision brackets are presented with the
following notations:
 \begin{eqnarray}
  \nonumber \widetilde K_1 &\equiv& \frac{K_3(z_k+z_l)}{K_2(z_k)K_2(z_l)},
  \quad \widetilde K_2\equiv \frac{K_2(z_k+z_l)}{K_2(z_k)K_2(z_l)},
  \quad \widetilde K_3\equiv \frac{G(z_k+z_l)}{K_2(z_k)K_2(z_l)}, \\
  Z_{kl} &\equiv& z_k+z_l, \quad z_{kl} \equiv z_k-z_l.
 \end{eqnarray}
For the scalar collision brackets one has:
 \eq{\label{br211}
 -[\tau,\tau_1]_{kl}=[\tau,\tau]_{kl}=\frac{\sigma^{cl}_{kl}}{\sigma(T)}\frac{\pi}{2z_k^2z_l^2Z_{kl}^2}
 (P_{s1}^{(1,1)}\widetilde K_1+P_{s2}^{(1,1)}\widetilde K_2+P_{s3}^{(1,1)}\widetilde K_3),
 }
where
 \eq{
 P_{s1}^{(1,1)}=-2 Z_{kl} (z_{kl}^4+4 z_{kl}^2 Z_{kl}^2-2 Z_{kl}^4),
 }
 \eq{
 P_{s2}^{(1,1)}=z_{kl}^4 (3 Z_{kl}^2+8)+32 z_{kl}^2 Z_{kl}^2+8 Z_{kl}^4,
 }
 \eq{
 P_{s3}^{(1,1)}=-3 z_{kl}^4 Z_{kl}^6,
 }
and
 \eq{
 [\tau,\tau_1^2]_{kl}=[\tau^2,\tau_1]_{lk}=\frac{\sigma^{cl}_{kl}}{\sigma(T)}\frac{\pi}{4z_k^2z_l^2Z_{kl}^2}
 (P_{s11}^{(1,2)}\widetilde K_1+P_{s12}^{(1,2)}\widetilde K_2+P_{s13}^{(1,2)}\widetilde K_3),
 }
where
 \eq{
 P_{s11}^{(1,2)}=2 Z_{kl} (z_{kl}^5 Z_{kl}+8 z_{kl}^4+16 z_{kl}^3 Z_{kl}
 +32 z_{kl}^2 Z_{kl}^2+16 z_{kl} Z_{kl}^3-40 Z_{kl}^4),
 }
 \begin{eqnarray}
  P_{s12}^{(1,2)}&=&-z_{kl}^5 Z_{kl} (Z_{kl}^2+8)-8 z_{kl}^4 (Z_{kl}^2+8)
  -16 z_{kl}^3 Z_{kl} (Z_{kl}^2+8)\\
  \nonumber &+&16 z_{kl}^2 Z_{kl}^2 (Z_{kl}^2-16)
  +8 z_{kl} Z_{kl}^3 (Z_{kl}^2-16)-8 Z_{kl}^4 (Z_{kl}^2+8),
 \end{eqnarray}
 \eq{
 P_{s13}^{(1,2)}=z_{kl}^5 Z_{kl}^7,
 }
and
 \eq{
 [\tau,\tau^2]_{kl}=[\tau^2,\tau]_{kl}=\frac{\sigma^{cl}_{kl}}{\sigma(T)}\frac{\pi}{4z_k^2z_l^2Z_{kl}^2}
 (P_{s21}^{(1,2)}\widetilde K_1+P_{s22}^{(1,2)}\widetilde K_2+P_{s23}^{(1,2)}\widetilde K_3),
 }
where
 \eq{
 P_{s21}^{(1,2)}=2 Z_{kl} (z_{kl}^5 Z_{kl}-8 z_{kl}^4+16 z_{kl}^3 Z_{kl}
 -32 z_{kl}^2 Z_{kl}^2+16 z_{kl} Z_{kl}^3+40 Z_{kl}^4),
 }
 \begin{eqnarray}
  P_{s22}^{(1,2)}&=&-z_{kl}^5 Z_{kl} (Z_{kl}^2+8)+
  8 z_{kl}^4 (Z_{kl}^2+8)-16 z_{kl}^3 Z_{kl} (Z_{kl}^2+8)\\
  \nonumber &-& 16 z_{kl}^2 Z_{kl}^2 (Z_{kl}^2-16)+8 z_{kl} Z_{kl}^3 (Z_{kl}^2-16)
  +8 Z_{kl}^4 (Z_{kl}^2+8),
 \end{eqnarray}
 \eq{
 P_{s23}^{(1,2)}=z_{kl}^5 Z_{kl}^7,
 }
and
 \eq{
 [\tau^2,\tau_1^2]_{kl}=\frac{\sigma^{cl}_{kl}}{\sigma(T)}\frac{\pi}{24z_k^2z_l^2Z_{kl}^2}
 (P_{s11}^{(2,2)}\widetilde K_1+P_{s12}^{(2,2)}\widetilde K_2+P_{s13}^{(2,2)}\widetilde K_3),
 }
where
 \begin{eqnarray}
  \nonumber P_{s11}^{(2,2)}&=&-2 Z_{kl} [z_{kl}^6 (Z_{kl}^2+2)+6 z_{kl}^4
  (11 Z_{kl}^2-32)-72 z_{kl}^2 Z_{kl}^2 (Z_{kl}^2+8)\\ &+&24 Z_{kl}^4 (Z_{kl}^2+96)],
 \end{eqnarray}
 \begin{eqnarray}
  P_{s12}^{(2,2)}&=&z_{kl}^6 (Z_{kl}^4+10 Z_{kl}^2+16)-6 z_{kl}^4
  (Z_{kl}^4-56 Z_{kl}^2+256)\\
  \nonumber &+& 144 z_{kl}^2 Z_{kl}^2 (5 Z_{kl}^2-32)
  -48 Z_{kl}^4 (13 Z_{kl}^2+32),
 \end{eqnarray}
 \eq{
 P_{s13}^{(2,2)}=-z_{kl}^4 Z_{kl}^6 [z_{kl}^2 (Z_{kl}^2-6)-6 Z_{kl}^2],
 }
and
 \eq{
 [\tau^2,\tau^2]_{kl}=\frac{\sigma^{cl}_{kl}}{\sigma(T)}\frac{\pi}{24z_k^2z_l^2Z_{kl}^2}
 (P_{s21}^{(2,2)}\widetilde K_1+P_{s22}^{(2,2)}\widetilde K_2+P_{s23}^{(2,2)}\widetilde K_3),
 }
where
 \begin{eqnarray}
  P_{s21}^{(2,2)}&=&-2 Z_{kl} [z_{kl}^6 (Z_{kl}^2+2)-36 z_{kl}^5
  Z_{kl}+18 z_{kl}^4 (Z_{kl}^2+16) \\
  \nonumber &+& 96 z_{kl}^3 Z_{kl} (Z_{kl}^2-10) + 24 z_{kl}^2 Z_{kl}^2 (Z_{kl}^2+56) \\
  \nonumber &-& 48 z_{kl} Z_{kl}^3 (Z_{kl}^2+20) - 24 Z_{kl}^4 (Z_{kl}^2+100)],
 \end{eqnarray}
 \begin{eqnarray}
  P_{s22}^{(2,2)}&=&z_{kl}^6 (Z_{kl}^4+10 Z_{kl}^2+16)+
  12 z_{kl}^5 Z_{kl} (Z_{kl}^2-24) \\
  \nonumber &-& 6 z_{kl}^4 (Z_{kl}^4-72 Z_{kl}^2-384) - 192 z_{kl}^3 Z_{kl} (Z_{kl}^2+40) \\
  \nonumber &-& 48 z_{kl}^2 Z_{kl}^2 (13 Z_{kl}^2-224) + 96 z_{kl} Z_{kl}^3 (7 Z_{kl}^2-80) \\
  \nonumber &+& 48 Z_{kl}^4 (13 Z_{kl}^2+48),
 \end{eqnarray}
 \eq{
 P_{s23}^{(2,2)}=-z_{kl}^4 Z_{kl}^6 [z_{kl}^2 (Z_{kl}^2-6)+12 z_{kl} Z_{kl}-6 Z_{kl}^2].
 }
And for the tensor collision brackets one has:
 \eq{
 [\overset{\circ}{\overline{\pi^{\mu} \pi^{\nu}}},
 \overset{\circ}{\overline{\pi_{1\mu} \pi_{1\nu}}}]_{kl}=
 \frac{\sigma^{cl}_{kl}}{\sigma(T)}\frac{\pi}{72z_k^2z_l^2Z_{kl}^2}(P_{T11}^{(0,0)}\widetilde K_1
 +P_{T12}^{(0,0)}\widetilde K_2+P_{T13}^{(0,0)}\widetilde K_3),
 }
where
 \begin{eqnarray}
  \nonumber P_{T11}^{(0,0)}&=&-2 Z_{kl} [z_{kl}^6 (5 Z_{kl}^2-8)+
  24 z_{kl}^4 (Z_{kl}^2-16)-144 z_{kl}^2 Z_{kl}^2 (Z_{kl}^2+8)\\
  &+&48 Z_{kl}^4 (Z_{kl}^2+72)],
 \end{eqnarray}
 \begin{eqnarray}
  P_{T12}^{(0,0)}&=&z_{kl}^6 (5 Z_{kl}^4-40 Z_{kl}^2-64)-
  24 z_{kl}^4 (5 Z_{kl}^4+8 Z_{kl}^2+128) \\
  \nonumber &+& 576 z_{kl}^2 Z_{kl}^2
  (Z_{kl}^2-16)-192 Z_{kl}^4 (5 Z_{kl}^2+16),
 \end{eqnarray}
 \eq{
 P_{T13}^{(0,0)}=-5 z_{kl}^4 Z_{kl}^6 [z_{kl}^2 (Z_{kl}^2-24)-24 Z_{kl}^2],
 }
and
 \eq{
 [\overset{\circ}{\overline{\pi^{\mu} \pi^{\nu}}},
 \overset{\circ}{\overline{\pi_{\mu} \pi_{\nu}}}]_{kl}=
 \frac{\sigma^{cl}_{kl}}{\sigma(T)}\frac{\pi}{72z_k^2z_l^2Z_{kl}^2}(P_{T21}^{(0,0)}\widetilde K_1
 +P_{T22}^{(0,0)}\widetilde K_2
 +P_{T23}^{(0,0)}\widetilde K_3),
 }
where
 \begin{eqnarray}
  \nonumber &~& P_{T21}^{(0,0)}=2 Z_{kl} [z_{kl}^6 (8-5 Z_{kl}^2)+72 z_{kl}^4 (3 Z_{kl}^2-8)
  -480 z_{kl}^3 Z_{kl} (Z_{kl}^2-4) \\
  &-& 336 z_{kl}^2 Z_{kl}^2 (Z_{kl}^2+8)+240 z_{kl} Z_{kl}^3
  (Z_{kl}^2+8)+192 Z_{kl}^4 (Z_{kl}^2+67)],
 \end{eqnarray}
 \begin{eqnarray}
  \nonumber P_{T22}^{(0,0)}&=&z_{kl}^6 (5 Z_{kl}^4-40 Z_{kl}^2-64)+240 z_{kl}^5
  Z_{kl}^3-24 z_{kl}^4 (5 Z_{kl}^4+48 Z_{kl}^2-192) \\
  \nonumber &+& 1920 z_{kl}^3 Z_{kl}
  (Z_{kl}^2-8)-192 z_{kl}^2 Z_{kl}^2 (17 Z_{kl}^2-112) \\
  &+& 1920 z_{kl} Z_{kl}^3 (Z_{kl}^2-8) + 768 Z_{kl}^4 (5 Z_{kl}^2+6),
 \end{eqnarray}
 \eq{
 P_{T23}^{(0,0)}=-5 z_{kl}^4 Z_{kl}^6 [z_{kl}^2 (Z_{kl}^2-24)+48 z_{kl} Z_{kl}-24 Z_{kl}^2].
 }
If ${z_k=z_l}$, then the $G(x)$ function is eliminated everywhere
and the collision brackets simplify considerably.

\end{document}